\documentclass[final,3p,times,authoryear]{elsarticle}
\usepackage{amssymb}
\usepackage{lipsum}
\usepackage{graphicx}
\usepackage{amsmath}
\usepackage{array}
\usepackage{subcaption}
\usepackage{xcolor}
\usepackage{url}
\usepackage{natbib}
\usepackage{hyperref}
\usepackage{array}
\usepackage{multirow}
\usepackage{multicol}
\usepackage{tablefootnote}
\usepackage{geometry}

\journal{Astronomy $\&$ Computing}
\begin{document}
\begin{frontmatter}



\title{Impact of cosmic web on galaxy properties and their correlations: Insights from Principal Component Analysis}


\author[first]{Anindita Nandi}
\ead{anindita.nandi96@gmail.com}
\author[first]{Biswajit Pandey}
\ead{biswap@visva-bharati.ac.in}

\affiliation[first]{organization={Department of Physics, Visva-Bharati University},%Department and Organization
            addressline={}, 
            city={Santiniketan},
            postcode={731235}, 
            state={West Bengal},
            country={India}}
\begin{abstract}
We use Principal Component Analysis (PCA) to analyze a volume-limited
sample from the SDSS and explore how cosmic web environments affect
the interrelations between various galaxy properties, such as $(u-r)$
colour, stellar mass, specific star formation rate, metallicity,
morphology, and $D4000$. Our analysis reveals that the first three
principal components (PC1, PC2 and PC3) account for approximately
$\sim 85\%$ of the data variance. We classify galaxies into different
cosmic web environments based on the eigenvalues of the deformation
tensor and compare PC1, PC2, PC3 across these environments, ensuring a
mass-matched sample of equal size for each environment. PC1 is
dominated by colour, sSFR, D4000, and morphology. It displays clear
bimodality across all cosmic web environments, with sheets and
clusters showing distinct preferences for negative and positive PC1
values, respectively. This variation reflects the strong role of
environmental processes in regulating star formation. PC2 and PC3,
respectively show positively and negatively skewed unimodal
distributions in all environments. PC2 is primarily influenced by
metallicity whereas PC3 is dominated by stellar mass. It indicates
that metallicity evolves gradually and is less sensitive to
environmental extremes, highlighting the importance of internal,
secular processes. PC3 likely captures residual variation in stellar
mass within the two main galaxy populations (star-forming and
quiescent) separated by PC1. A Kolmogorov-Smirnov (KS) test confirms
that the distributions of PC1, PC2 and PC3 differ significantly across
environments, with a confidence level exceeding
$99.99\%$. Furthermore, we calculate the normalized mutual information
(NMI) between the principal components and individual galaxy
properties within different cosmic web environments. A two-tailed
t-test reveals that for each relationship and each pair of
environments, the null hypothesis is rejected with a confidence level
$>99.99\%$. Our analysis confirms that cosmic web environments play a
significant role in shaping the correlations between galaxy
properties.
\end{abstract}



\begin{keyword}
%% keywords here, in the form: keyword \sep keyword, up to a maximum of 6 keywords
% keyword 1 \sep keyword 2 \sep keyword 3 \sep keyword 4
cosmology: large-scale structure of Universe; 
galaxies: statistics;
methods: data analysis


\end{keyword}

\end{frontmatter}

%\tableofcontents



\section{Introduction}
\label{introduction}

The origin and evolution of galaxies stands as a paramount challenge
in modern cosmology. It remains at the forefront of research in
cosmology over recent decades. According to the current paradigm, the
first bound structures emerge from the collapse of initial
fluctuations into dark matter halos. These halos accumulate neutral
hydrogen gas from their surroundings via accretion, which eventually
condenses and cools, forming galaxies at their cores
\citep{reesostriker77, silk77, white78, fall80}. In this narrative,
the primary mechanism driving galaxy growth is the accretion of gas
from the intergalactic medium (IGM). As galaxies evolve, the formation
of a supermassive black hole at their centers and the efficient
accretion onto it can ignite active galactic nuclei (AGN)
activity. The feedback from the energetic processes like AGN,
supernovae or shock-driven winds can drive gas outflows from
galaxies. The expelled gas cools over time and reenters the galaxy,
perpetuating the cycle.

The gas inflow and outflow can modulate the galaxy properties. Gas
inflow supplies the raw fuel for star formation. It can trigger bursts
of star formation, influencing the mass, colour and morphology of
galaxies. The inflowing gas also brings pristine material into
galaxies, which affects their chemical composition. Conversely,
outflows can regulate star formation by ejecting gas from the galaxy,
thereby reducing the available gas reservoir. Strong outflows can even
lead to the quenching of star formation in galaxies. Moreover,
outflows transport the heavy elements synthesized in stars into the
interstellar medium (ISM) and IGM leading to metal enrichment of the
gas over cosmic time.

It is also crucial to remember that galaxies do not evolve in
isolation. The interaction of galaxies with their environment plays a
very important role in their evolution. The influence of environment
on shaping galaxy properties has been thoroughly explored in the
literature \citep{davis2, dress, butcher84, guzo, zevi, goto, hog1,
  blan1, einas2, balogh02, kauffmann04, mandel06, abbas06, pandey07,
  mouhcine, bamford, cooper10, koyama}. Numerous studies using
simulations \citep{toomre72,barnes96, mihos96, tissera02, cox06,
  montuori10, lotz11, torrey12, hopkins13, renaud14, renaud15,
  moreno15, moreno21, renaud22, das24} and observations
\citep{larson78, barton00, lambas03, alonso04, nikolic04, woods06,
  woods07, barton07, ellison08, ellison10, woods10, patton11,
  barrera15, thorp22, shah22, das22} have confirmed that tidal torques
resulting from galaxy interactions can trigger starbursts and modify
the colour and morphology of galaxies. Further, feedback mechanisms
are not the sole drivers of gas outflows and quenching in
galaxies. Several environment driven mechanisms such as
merger\citep{hopkins08}, harassment \citep{moore96, moore98},
strangulation \citep{gunn72, balogh00}, starvation \citep{larson80,
  somerville99, kawata08} and satellite quenching \citep{geha12} can
shut down star formation in galaxies and alter their
structures. Additional routes that can halt star formation in galaxies
include mass quenching \citep{birnboim03, dekel06, keres05, gabor10},
morphological quenching \citep{martig09}, bar quenching
\citep{masters10} and angular momentum quenching \citep{peng20}.

The environment of a galaxy is primarily characterized by the local
density. Most earlier studies investigating the impact of environment
on galaxy evolution have utilized local density as a proxy measure for
assessing environmental influences. However, galaxies are organized
into a complex interconnected structure of filaments, clusters,
sheets, and voids known as the cosmic web \citep{gregory78, joeveer78,
  einasto80, zeldovich82, einasto84, bond96, bharadfil, pandeyfil,
  arag10b, libeskind18}, indicating that their environment cannot be
fully characterized by local density alone. The cosmic web acts as a
scaffolding that organizes and shapes the distribution of matter and
galaxies on cosmic scales. Studies with N-body simulations reveal a
dynamic flow of matter through the cosmic web: from voids to walls,
walls to filaments, and ultimately into clusters \citep{arag10a,
  cautun14, ramachandra, wang24}. Moreover, hydrodynamical simulations
suggest that more than $(40-50)\%$ baryonic matter resides in
filaments as the Warm-Hot Intergalactic Medium (WHIM) which can
profoundly impact the efficiency of gas accretion onto galaxies
\citep{tuominen21, galarraga21}. The galaxies residing in different
cosmic web environment may have different gas accretion efficiency
\citep{cornu18, zhu22}. Galaxies located near the centres of cosmic
filaments and sheets benefit from a consistent influx of cold gas,
fueling vigorous star formation and boosting their overall mass
\citep{chen15, pandey20, singh20, laigle21, das22, das23, hoosain24}.
A recent study \citep{bulichi24} using the SIMBA simulation
\citep{dave19} suggests that shock heating in filaments can also
suppress star formation. Another study \citep{hasan24} utilizing the
IllustrisTNG simulation \citep{nelson19} indicates that satellite
galaxies are significantly more susceptible to such quenching within
the filaments. Clusters are the most densely populated regions within
the cosmic web, typically forming at the intersections of
filaments. The galaxies in such environment experience heightened gas
accretion rates and increased interactions with neighboring galaxies,
the cluster potential, and the gas \citep{treu03, delucia12}, often
leading to intense bursts of star formation and structural
changes. Recent studies suggest that galaxies located nearer to
clusters show higher gas-phase metallicity independent of stellar mass
and overdensity, indicating enhanced chemical enrichment compared to
galaxies farther away \citep{donnan22}. In contrast, galaxies residing
in the sparser regions and voids tend to undergo quieter evolutionary
paths, marked by more subdued star formation activities
\citep{maret22, rodriguez24}. This intricate interplay within the
cosmic web underscores its pivotal role in shaping the diverse
trajectories of galaxy evolution across the universe.

Several studies \citep{zehavi11, yan13, alpaslan15, alam19} do not
detect a significant impact of tidal environments on galaxy
properties, attributing observed differences to variations in the
underlying halo mass function within the cosmic web or the assembly
history of dark matter halos. In recent years, observational evidence
has increasingly shown that galaxy properties are also influenced by
their large-scale environment \citep{pandey2, pandey3, trujillo,
  erdogdu, paz, jones, scudder, tempel1, tempel2, darvish, filho,
  lupa, alpaslan16, pandey17, kuutma17, chen17, lee18, laigle18,
  kraljic18, chen19, pandey20, kraljic20, bonjean21, winkel21,
  malavasi22, bhambani23}. Galaxy properties such as stellar mass,
colour, morphology, star formation rate, star formation history, and
metallicity are correlated with each other, and these connections
might change depending on the cosmic web environment. The influence of
various morphological environments within the cosmic web on shaping
the correlations among galaxy properties remains relatively unclear. A
recent study \citep{nandi23} use Pearson Correlation Coefficient (PCC)
and Normalized Mutual Information (NMI) to analyze the correlations
between galaxy properties in different environments of the cosmic
web. The study demonstrates that the scaling relations between
observable galaxy properties are influenced by the geometric
environments of the cosmic web. In the present work, we will use
Principal Component Analysis (PCA) \citep{pearson} to analyze data
from the Sloan Digital Sky Survey (SDSS). The Sloan Digital Sky Survey
(SDSS) \citep{stout02} stands as one of the largest galaxy redshift
surveys conducted to date. The availability of precise photometric and
spectroscopic information for a large number of galaxies in the SDSS
has enabled accurate measurement of their physical properties. This
rich data provides an excellent opportunity for analyzing the impact
of the cosmic web on the correlations between various galaxy
properties using PCA.

PCA is a powerful statistical technique for analyzing complex datasets
and extracting meaningful patterns. Some earlier studies use Principal
Component Analysis (PCA) to explore complex relationships among
multiple variables in cosmological datasets. For instance,
\cite{einasto11} employ PCA to examine the properties of SDSS DR7
superclusters. \cite{chaves20} utilize PCA to investigate how the
intricate interplay of star formation history, metallicity evolution,
and dust properties influence present-day galaxy
colours. \cite{balaguera24} apply PCA to analyze the sensitivity of
scaling relations between various halo properties with respect to
their environment. PCA offers several advantages over the traditional
statistical measures for studying correlations between
variables. Galaxy properties are often characterized by a large number
of variables, making it challenging to interpret their
relationships. By transforming the original variables into a smaller
set of orthogonal components, PCA can reveal the underlying structure
of relationships.  PCA effectively captures the maximum variance in
the data, allowing us to focus on the most significant relationships
while discarding noise. Further, it can reveal emergent behaviours not
immediately apparent from studying individual properties, as the
components reflect combined contributions. We will carry out a
comparative analysis of the principal components and their
relationships with different galaxy properties across different cosmic
web environments. This could provide valuable insights into the
significance of the physical processes that influence galaxy evolution
within distinct cosmic web environments.

The paper is structured as follows: Section 2 describes the data,
Section 3 outlines the analysis method, Section 4 discusses the
results, and Section 5 presents the conclusions.

In this study, we adopt a $\Lambda$CDM cosmology with
$\Omega_{m}=0.315$, $\Omega_{\Lambda}=0.685$, and $h=0.674$
\citep{planck18}. These parameters are used to convert redshifts to
comoving distances throughout the analysis.

%%%%%%%%%%%%%%%%%%%%%%%%%%%%%%%%%%%%%%%%%%%%%%%%%%%%%%%%%%%%%%%%%%%%%%%%%%%%%%%%%%%
\section{Data}
\label{sec:data}
The SDSS \citep{stout02} is one of the most ambitious and influential
redshift surveys todate. It uses a dedicated 2.5-meter telescope at
Apache Point Observatory in New Mexico. SDSS has measured the
photometric and spectroscopic information of millions of galaxies,
stars, and quasars across a significant portion of the sky. SDSS DR18
\citep{almeida} is the eighteenth data release of the SDSS. We
download the data using \textit{Structured Query Language} (SQL) from
\textit{CasJobs} \footnote{\url{https://skyserver.sdss.org/casjobs/}}. We
extract spectroscopic and photometric data pertaining to galaxies from
the \textit{SpecObj}, \textit{PhotoObj}, and \textit{Photoz}
tables. We restrict our search to galaxies with the
\textit{scienceprimary} flag set to 1, ensuring inclusion of those
with the highest-quality spectroscopic data. In our analysis, we
utilize observed colours that have not been corrected for reddening
caused by redshift or internal extinction.  We quantify morphology of
galaxies using the concentration index $\frac{r_{90}}{r_{50}}$
\citep{shimasaku01}, where $r_{90}$ and $r_{50}$ represent the radii
containing $90\%$ and $50\%$ of the Petrosian flux, respectively. The
$r_{90}$ and $r_{50}$ were retrieved from the \textit{PhotoObj}
table. The stellar mass, specific star formation rate (sSFR) and
metallicity of the galaxies are derived from the table
\textit{StellarMassFSPSGranWideDust} \citep{conroy09}.  sSFR
represents the star formation rate per unit galaxy stellar mass. The
metallicity refers to the proportion of elements heavier than Helium
\citep{asplund09}, estimated using the Flexible Stellar Population
Synthesis (FSPS) technique \citep{conroy09}. These properties are
determined by comparing observed spectroscopic and/or photometric
properties of galaxies with stellar population synthesis models. SDSS
spectra are obtained through a $3$ arcsec aperture covering only a
fraction of the entire galaxy. One can apply corrections for the
finite aperture of the SDSS fibres \citep{brinchmann04}. However,
\cite{conroy09} demonstrate that relying solely on broadband
photometry yields more robust results. We obtain the strength of the
$4000\,\mathring{A}$ break (D4000) \citep{bruzual83, balogh99} from
the \textit{galSpecIndx} table. D4000 characterizes the mean age of
the stellar population in the galaxy.

\begin{figure}
\centering 
\begin{minipage}{0.48\textwidth}
    \centering
\includegraphics[width=\linewidth]{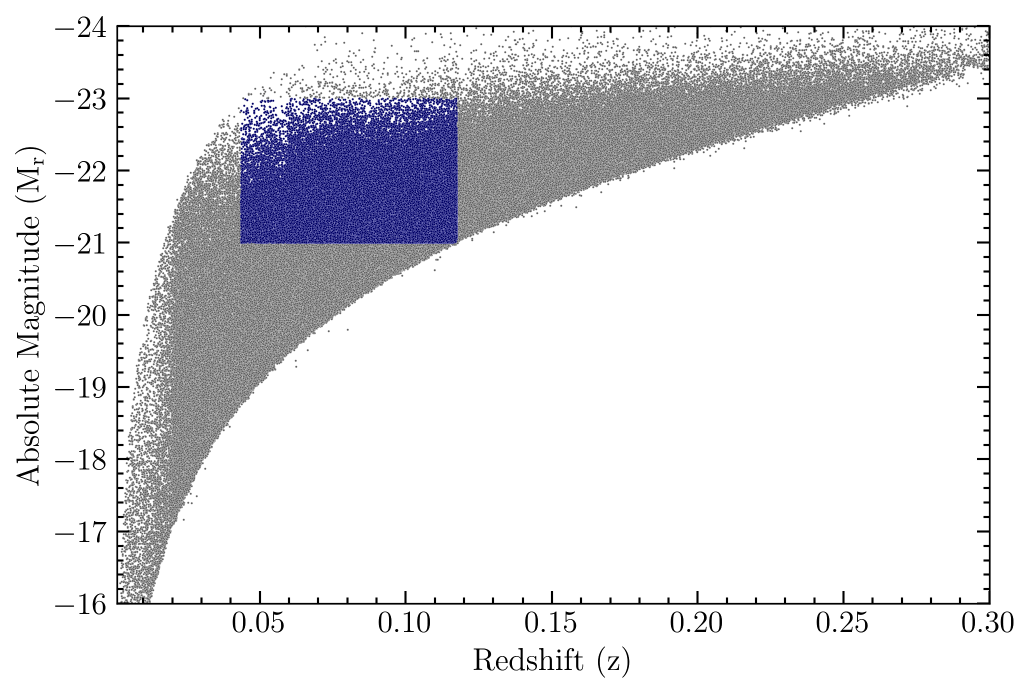}
\end{minipage}
\hfill
\begin{minipage}{0.48\textwidth}
    \centering
    \includegraphics[width=\linewidth]{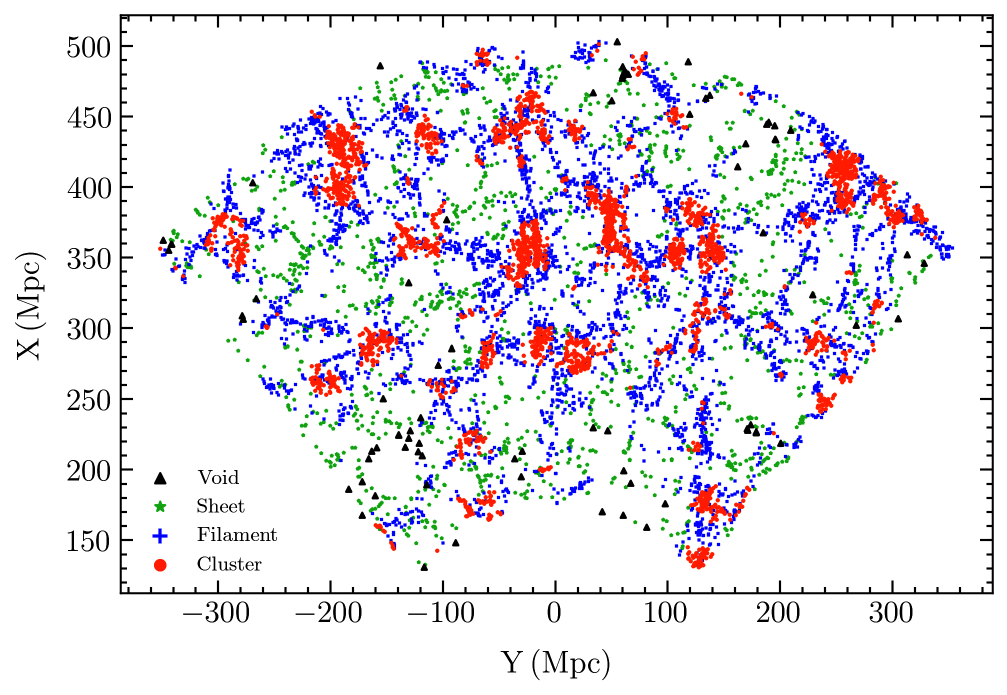}
\end{minipage}

\caption{The left panel of this figure shows the distribution of SDSS
  DR18 galaxies in the redshift-absolute magnitude plane. The blue
  region in this diagram represents our volume-limited sample. The
  right panel displays a projected distribution of a $30$ Mpc thick
  slice of this volume-limited sample in the X-Y plane. Different
  geometric environments of the cosmic web, identified using the
  Hessian-based method, are highlighted in various colours.}
\label{fig:vsample}	
\end{figure}

We construct a volume limited galaxy sample using the SDSS DR18
data. We identify a contiguous region in the sky between the right
ascension $135^{\circ}\leq \alpha \leq 225^{\circ}$ and declination
$0^{\circ} \leq \delta \leq 60^{\circ}$.  The volume limited sample is
extracted by applying extinction-corrected and $k$-corrected $r$-band
absolute magnitude cut $-23 \leq M_r \leq -21$, which corresponds to
the redshift range $0.0434 \leq z \leq 0.1175$
(\autoref{fig:vsample}).  We consider specific ranges of galaxy
properties (Table \autoref{tab:galprops_range}) for the present
analysis, excluding any extreme outliers. Finally, we obtain $88579$
galaxies in our volume limited sample.

\renewcommand{\arraystretch}{1.5} % Increase row height by 1.5 times

\begin{table}[htbp]
    \centering
    \begin{tabular}{|c|c|c|}
    \hline
       Galaxy property  & Minimum & Maximum \\
       \hline
       \hline
        $(u-r)\,colour$ & $0.5$ & $4.5$ \\
        \hline
        $\log_{10}\left(\frac{M_{\star}}{M_{\odot}}\right)$ & $10.0$ & $12.0$ \\
        \hline
        $\log_{10}(sSFR/Gyr^{-1})$ & $-22$ & $-0.2$ \\
        \hline
        Metallicity & $0.007$ & $0.03$ \\
        \hline
        $\frac{r_{90}}{r_{50}}$ & $1.5$ & $4.5$ \\
        \hline
        $D4000$ & $0.8$ & $2.5$ \\
        \hline
    \end{tabular}
    \caption{This table shows the range of different galaxy properties in our final sample.}
    \label{tab:galprops_range}
\end{table}
% ##################################################################### %
% ##################################################################### %

%%%%%%%%%%%%%%%%%%%%%%%%%%%%%%%%%%%%%%%%%%%%%%%%%%%%%%%%%%%%%%%%%%%%%%%%%%%%%%%%%%%

%%%%%%%%%%%%%%%%%%%%%%%%%%%%%%%%%%%%%%%%%%%%%%%%%%%%%%%%%%%%%%%%%%%%%%%%%%%%%%%%%%%
\section{Method}

\subsection{Classifying different morphological environments in the Cosmic Web}\label{sec:method2}

We use a Hessian based classification of the cosmic web \citep{hahn2,
  fromero} for the present work. This method classifies the different
morphological environments of the cosmic web based on the eigenvalues
and eigenvectors of the deformation tensor.

The deformation tensor $T_{ij}$ is described by the Hessian matrix of
the gravitational potential field $\Phi$ as,
\begin{equation}
T_{ij} = \frac{\partial^2 \Phi}{\partial x_i \partial x_j}, 
\end{equation}
where $x_i$ and $x_j$ are the spatial coordinates.

We compute gravitational potential $\Phi$ by solving the Poisson
equation,
\begin{equation}
\nabla^2 \Phi \equiv \delta
\end{equation}
where $\delta = \frac{\rho-\bar{\rho}}{\rho}$ is the density
contrast. We first apply the Cloud-In-Cell (CIC) scheme on a $256^3$
grid to construct a discrete density contrast field. The overdensity
field is then Fourier transformed and smoothed using an isotropic
Gaussian filter of width $8\,\mathrm{Mpc}$, which is close to the mean
intergalactic separation of our galaxy sample $\approx 8.57\,\rm Mpc$.
This choice restricts our ability to characterize environments on
scales smaller than the intergalactic separation. However, our
approach focuses on quantifying the large-scale geometric environments
within the cosmic web.

We compute the Fourier transform of the gravitational potential
associated with fluctuations in the smoothed density field,
\begin{equation}
\hat{\Phi} = \hat{\mathcal{G}}\hat{\rho}
\end{equation}
In this approach, $\hat{\mathcal{G}}$ represents the Fourier transform
of the Green's function of the Laplacian operator, and $\hat{\rho}$
denotes the density in Fourier space. We transform the potential back
into real space and then compute the tidal tensor using numerical
differentiation. Based on the signs of its three eigenvalues
($\lambda_1$, $\lambda_2$, $\lambda_3$), we categorize each galaxy
into voids, sheets, filaments, or clusters. Each galaxy is
classified as a part of
\begin{enumerate}
    \item void , if $\lambda_1,\lambda_2,\lambda_3 < 0$
    \item sheet , if $\lambda_1 > 0,\, \lambda_2,\lambda_3 < 0$
    \item filament , if $\lambda_1,\lambda_2 > 0,\, \lambda_3 < 0$
    \item cluster , if $\lambda_1,\lambda_2 , \lambda_3 > 0$
\end{enumerate}
where $\lambda_1 > \lambda_2 > \lambda_3$. The total number of
galaxies in our volume limited sample, residing in voids, sheets,
filaments and clusters are listed in Table~\ref{tab:galaxies_in_web}.

\begin{table}[htbp]
    \centering
    \begin{tabular}{|c|c|}
    \hline
     Cosmic Web Environment & Number of Galaxies \\
     \hline
     \hline
       Void  & $1273$ \\
       \hline
       Sheet & $14597$ \\
       \hline
       Filament & $45028$ \\
       \hline
       Cluster & $27681$ \\
       \hline
    \end{tabular}
    \caption{This table provides the number of galaxies identified in
      different cosmic web environments.}
    \label{tab:galaxies_in_web}
\end{table}
% ********************************************************************* %
\begin{figure*}[htbp]
\centering
\includegraphics[width=0.9\textwidth]{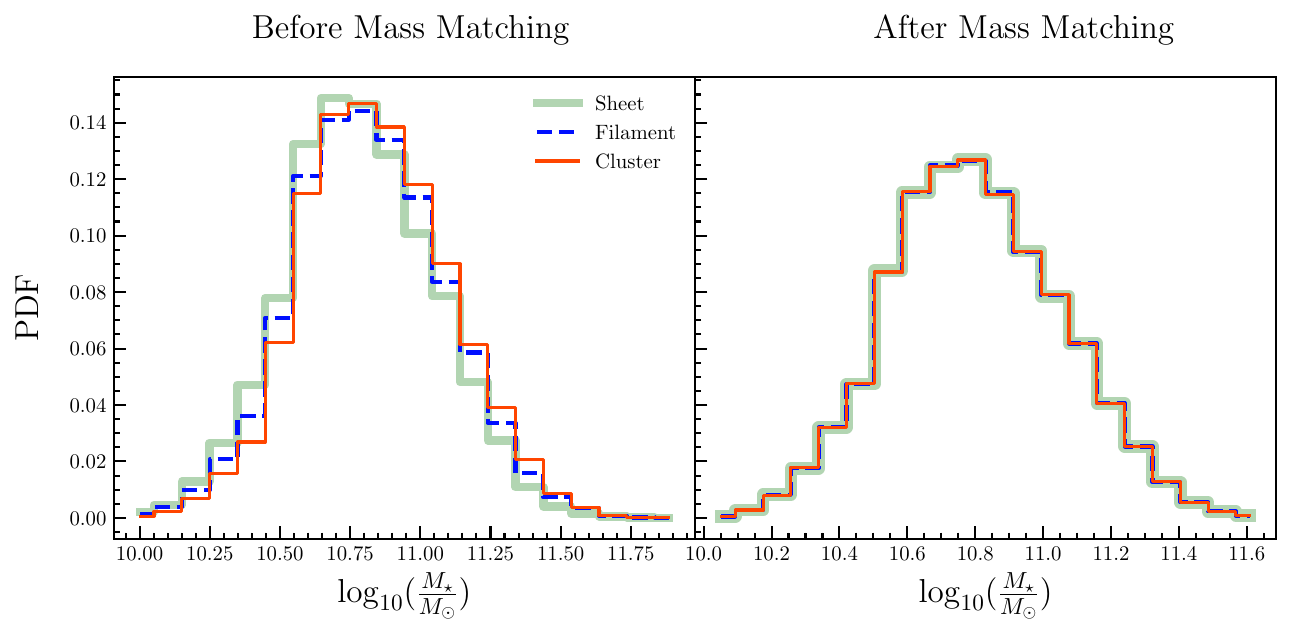}
\caption{The left panel of this figure compares the PDFs of
  $\log_{10}(\frac{M_{\star}}{M_{\odot}})$ for the galaxies in
  different cosmic web environments. The right panel shows the same
  but after matching the stellar mass of galaxies in different
  environments.}
\label{fig:web}
\end{figure*}
%%%%%%%%%%%%%%%%%%%%%%%%%%%%%%%%%%%%%%%%%%%%%%%%%%%%%%%%%%%%%%%%%%%%%%%%%%%%%%%%%%%%%%%%%%%%%
We utilize the entire volume-limited sample consisting of $88579$
galaxies to identify the various morphological components of the
cosmic web.

Galaxy properties are strongly influenced by mass, necessitating
careful consideration to avoid bias in any correlation
studies. Differences in correlations across various geometric
environments may stem from varied stellar mass distributions in
different regions of the cosmic web. We mitigate this issue by
matching the stellar mass distribution in different types of
environments. We ensure that an equal number of galaxies from each of
the three geometric environments are matched in terms of their stellar
masses, with a precision of $0.001$ dex. We show the PDF of
$\log_{10}(\frac{M_{\star}}{M_{\odot}})$ for galaxies in different
cosmic web environments in the left and right panels of
Figure~\ref{fig:web} before and after the mass-matching
respectively. The stellar-mass matching yeields $14339$ galaxies each
in sheets, filaments and clusters. It may be noted that the final
number of matched galaxies in other environments are not exactly equal
the number of galaxies present in sheets
(Table~\ref{tab:galaxies_in_web}). This difference arises because not
every galaxy in the reference sample (sheet in this case) meets our
mass matching criteria. We do not consider the voids due to the very
small number galaxies present within them. After the mass-matching, we
perform a KS test to assess the statistical significance of the
differences in the stellar mass distribution of galaxies from sheets,
filaments and clusters. We observe that the null hypothesis can be
confidently accepted at a significance level exceeding $99.99\%$ for
each pair of environment. The test confirms that the stellar mass
distributions in our mass-matched sample are statistically
indistinguishable across different environments. The mass-matched
sample finally consists of $43017$ galaxies from three different
cosmic web environments. The analysis throughout the rest of the paper
is based on this stellar mass-matched sample.

%%%%%%%%%%%%%%%%%%%%%%%%%%%%%%%%%%%%%%%%%%%%%%%%%%%%%%%%%%%%%%%%%%%%%%%%%%%%%%%%%%%%%%%%%%%%%
\subsection{Principal Component Analysis (PCA)}\label{sec:method1}

PCA \citep{pearson} is a statistical technique used to reduce the
dimensionality of data while retaining as much variance as possible by
transforming the original variables into a new set of uncorrelated
variables called principal components. Here, we briefly describe the
steps involved in our analysis.

We consider a dataset $X$ consisting of $n$ observations (rows) and
$p$ variables (columns). In this analysis, our dataset $X$ comprises
$43017$ observations of $6$ variables. These observations represent
all galaxies in our mass-matched volume limited sample, for which we
gathered data on six distinct physical properties: colour, stellar
mass, specific star formation rate (sSFR), metallicity, concentration
index, and the D4000.

First, we center the data $X$ by subtracting the mean from each
column, and then divide each element by the standard deviation of the
respective column as,

\begin{equation}
X^{\prime}_{ij} = \frac{X_{ij} - \mu_j}{\sigma_j}, 
\end{equation}
where $\mu_j = \frac{1}{n} \sum_{i=1}^{n} X_{ij}$ is the mean and
$\sigma_j = \sqrt{\frac{1}{n-1} \sum_{i=1}^{n} (X_{ij} - \mu_j)^2}$ is
the standard deviation for each column. Here $i$ corresponds to rows
($n$ galaxies) and $j$ corresponds to columns ($p$ physical
properties).
  
We then compute the covariance matrix of the standardized data as,
\begin{equation}
 S = \frac{1}{n-1} X^{\prime T} X^{\prime},
\end{equation}
where $X^{\prime T}$ denotes the transpose of $X^{\prime}$. Here, $S$
is the covariance matrix among the variables of the data matrix and it
has a size of $p \times p$. We determine the eigenvectors and the
eigenvalues of the covariance matrix $S$.

PCA aims to find a set of orthogonal vectors (principal components)
that capture the maximum variance in the data. These principal
components correspond to the eigenvectors of the covariance matrix
$S$. Let $v_1, v_2,....,v_p$ be the eigenvectors of $S$, and
$(\lambda_1, \lambda_2,...., \lambda_p)$ be the corresponding
eigenvalues. The eigenvalues represent the variance explained by each
eigenvector. We order the principal components by the amount of
variance they explain. This simply corresponds to the magnitude of
their eigenvalues. The first principal component (PC1) captures the
most variation in the data and the subsequent components (PC2,
PC3,...) explain the remaining variation in descending order of
importance.

The eigenvectors derived from the covariance matrix can be arranged
into a matrix, where each column corresponds to an eigenvector.  We
multiply the standardized data matrix with this eigenvector
matrix. Each column of the resulting matrix represents a principal
component. Thus the $i^{th}$ principal component is given by PC$_i =
X^{\prime}\,v_i$, where $X^{\prime}$ is the standardized data matrix
and $v_i$ is the $i^{th}$ eigenvector.  Each principal component is a
linear combination of the original variables, weighted by the elements
of the corresponding eigenvector.  The weight or the coefficient of
any particular variable in this combination tells us the importance of
that variable in defining that component. We analyze how individual
galaxy properties relate to specific principal components across
different cosmic web environments. This may provide valuable insights
into the physical processes governing galaxy evolution in different
geometric environments. This would also enable us to explore how
different galaxy properties relate to each other in terms of variance
captured by the principal components in different environments.

\subsection{Normalized Mutual Information  (NMI)}\label{sec:method3}
Mutual information (MI) is a non-parametric measure from information
theory capable of identifying various forms of relationships,
including those that are non-linear and non-monotonic. MI does not
rely on any specific assumptions regarding the
distributions. Consequently, it is generally regarded as a more
general and robust measure of association compared to the Pearson or
Spearman correlation coefficient \citep{spearman}.

Consider a discrete random variable $X$ with $n$ possible outcomes,
denoted as \{$X_i : i = 1,\ldots n$\}. Let $P(X_i)$ represent the
probability of the $i^{th}$ outcome. The information entropy
\citep{shannon48} for $X$ is defined as,
\begin{equation}
H(X) = -\sum_{i=1}^{n} P(X_i)\,\log P(X_i).
\end{equation} 
Here, we choose the base of the logarithm to be $10$.

Now, let $X$ and $Y$ be two discrete random variables representing
different galaxy properties. The joint entropy of $X$ and $Y$ is given
by,
\begin{equation}
H(X,Y) = -\sum_{i=1}^{n_1}\sum_{j=1}^{n_2} P(X_i,Y_j)\,\log P(X_i,Y_j).
\end{equation}
In our analysis, we use $ n_1 = n_2 = 20$. The joint probabilities
$P(X_i,Y_j)$ for different values of $X_i$ and $Y_j$ are computed from
the joint distribution of the two random variables. These joint
probabilities represent the normalized abundance of galaxies within
specific ranges of X and Y.

MI between $X$ and $Y$ is calculated as,
\begin{equation}
I(X;Y) = H(X) + H(Y) - H(X,Y)
\end{equation}

MI measures how much information one variable provides about another,
irrespective of the nature of their relationship. It does not assume a
linear relationship between $X$ and $Y$. When $X$ and $Y$ are
statistically independent, the mutual information is zero.

The normalized mutual information (NMI) \citep{strehl02} is a
normalized measure of mutual information, defined as,
\begin{equation}
NMI (X;Y) = \frac{I(X;Y)}{\sqrt{H(X)H(Y)}}
\end{equation}
The NMI value spans from 0 to 1, where a value of 1 signifies a strong
association between the variables, and a value of 0 indicates no
association.

Entropy can be sensitive to the choice of binning, but this issue is
mitigated if comparisons are made using a consistent number of
bins. In our analysis, we maintain the same number of bins and the
same number of galaxies for each type of geometric environment. This
approach ensures that shot noise contributes equally across different
environments, allowing for meaningful comparisons of results.

In this study, we will analyze the NMI between the principal
components and galaxy properties across various geometric
environments. Our goal is to determine whether these relationships
differ in a statistically significant way.

\begin{figure*}[htbp]
\centering
\includegraphics[width=1.0\textwidth]{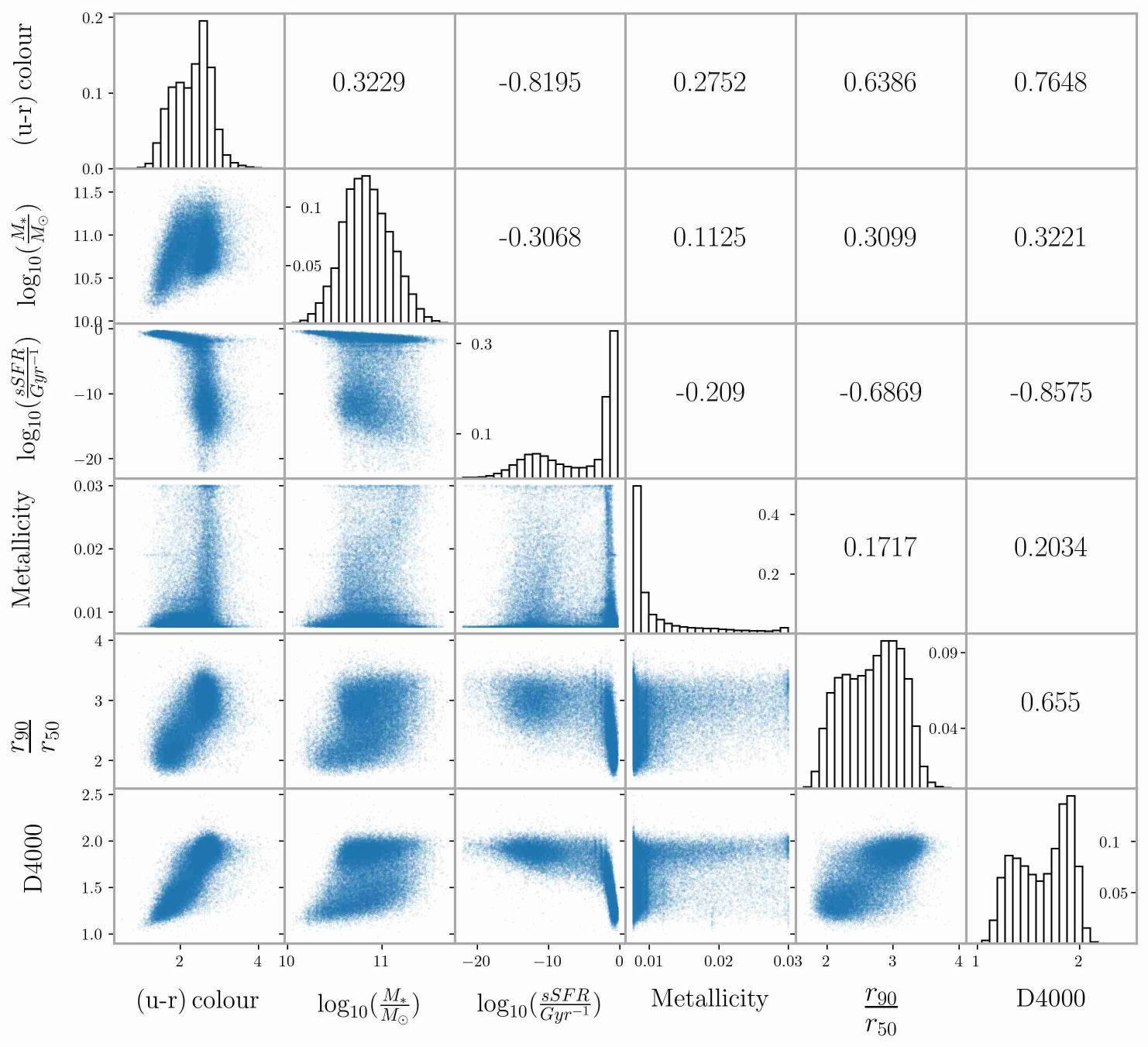}
\caption{The lower diagonal panels of this plot show the scatter plots
  for the different pairs of galaxy properties. The panels in the
  upper diagonal show the Spearman's rank correlation coefficients for
  the respective pairs of galaxy properties. The diagonal panels show
  the PDF for individual galaxy properties.}
\label{fig:prop_corrs}
\end{figure*}

%%%%%%%%%%%%%%%%%%%%%%%%%%%%%%%%%%%%%%%%%%%%%%%%%%%%%%%%%%%%%%%%%%%%%%%%%%%%%%%%%%%%%%%%%%
\section{Results}
\label{sec:results}

\subsection{Analyzing the correlations between different galaxy properties with Spearman's rank correlation}
\label{subsec:spearman}
Different galaxy properties are interrelated with each other. We first
study the correlations between various galaxy properties in our
combined sample using Spearman's rank correlation coefficient. The
Spearman's rank correlation coefficient is a nonparametric measure
that can quantify the statistical dependence between the rankings of
two variables, gauging the extent to which their relationship can be
characterized by a monotonic function. In other words, it evaluates
whether the variables tend to change together in the same direction
without assuming they change in a linear fashion. Spearman's rank
correlation is thus suitable for assessing relationships that are
monotonic but not necessarily linear. This allows for a more
comprehensive data analysis, capturing nonlinear associations that may
exist between different galaxy properties.  The Spearman's rank
correlation coefficient ranges from $-1$ to $1$. If the value of the
correlation coefficient is 1, it indicates a perfect positive
correlation between two variables, implying that as one variable
increases, the other variable increases proportionally. Conversely, a
value of -1 signifies a perfect negative correlation and a coefficient
of 0 suggests no correlation between the variables.

We use Spearman's rank correlation to quantify the strength and
direction of the relationships. This analysis tells us about the
existing relationships between different galaxy properties in our
sample. We consider the entire volume limited galaxy sample without
partitioning it into distinct geometric environments. The values of
Spearman's rank correlation coefficient as obtained in our analysis,
are shown in \autoref{fig:prop_corrs}. We observe the highest positive
correlation $(\sim 0.7648)$ between $(u-r)$ colour and $D4000$ and the
highest negative correlation $(\sim -0.8575)$ between $D4000$ and
$\log(sSFR)$. The correlation values between metallicity and other
properties are notably low. This is also corroborated by the scatter
plots in \autoref{fig:prop_corrs}. We can observe considerable
dispersion between metallicity and other galaxy properties.

This analysis provides us with a summary of the strength and direction
of the relationships between different galaxy properties in our
sample. It is clear that various galaxy properties exhibit both linear
and non-linear relationships. Identifying and understanding the nature
of both linear and non-linear relationships among multiple variables
can be quite complex. Our primary goal in this study is to verify if
the interrelationships between different galaxy properties are
affected by their large-scale cosmic web environment. PCA is a
powerful exploratory tool that can provide valuable insights into the
structure of data and relationships between multiple variables
exhibiting complex patterns. In the next subsection, we utilize PCA to
investigate the structure and relationships among galaxy properties in
our volume-limited sample. Additionally, we compare both the principal
components and their individual relationships with different galaxy
properties across various geometric environments within the cosmic
web.

\subsection{Understanding structure and relationships in the data using Principal Component Analysis}
\label{subsec:pca_data}
Our data matrix consists of $(u-r)$ colour,
$\log_{10}(\frac{M_{\star}}{M_{\odot}})$, $\log_{10}(sSFR)$,
metallicity, $\frac{r_{90}}{r_{50}}$, $D4000$ for $43017$ galaxies in
our volume limited sample. We first standardize the data matrix
following the method described in \autoref{sec:method1}. Subsequently,
we calculate the covariance matrix and determine its eigenvalues and
eigenvectors. The elements of the eigenvectors and the associated
galaxy properties are described in Table~\ref{tab:pca}. PCA converts
the set of galaxy properties to a new set of uncorrelated components,
known as the principal components (\autoref{sec:method1}). Each
principal component is a linear combination of the six galaxy
properties. The coefficients of the linear transformation determine
the contribution of each property to a particular principal
component. The largest variance of the data set lies along the first
principal component, while subsequent components capture decreasing
amounts of variance.

After performing PCA, we obtain a set of eigenvalues, each
corresponding to a principal component and representing the amount of
variance explained by that component. The percentage of explained
variance is calculated by dividing each eigenvalue by the sum of all
the eigenvalues and then multiplying by 100. This straightforward
calculation allows us to quantify how much variance each component
accounts for in the overall dataset. We show the percentage of
variance captured by the different principal components in
\autoref{tab:variance}. Approximately $85\%$ of the variance in our
data matrix is explained by the first three principal components. With
this in mind, our focus shifts to examining the behaviors of PC1 PC2,
and PC3 across different cosmic web environments. Additionally, we
investigate how principal components PC1, PC2 and PC3 relate to
various galaxy properties within these diverse geometric
environments. We describe the results of these analysis in the
following subsection.

%%%%%%%%%%%%%%%%%%%%%%%%%%%%%%%%%%%%%%%%%%%%%%%%%%%%%%%%%%%%%%%%%%%%%%%%%%%%%%%%%%%
\begin{table*}[htbp]
\centering
\begin{tabular}{|c|c|c|c|c|c|c|c}
\hline
Properties & $v_1$ & $v_2$ & $v_3$ & $v_4$ & $v_5$ & $v_6$\\
\hline
\hline
$(u-r)\,$ colour & $0.4836$ & $-0.0686$ & $-0.0192$ & $ 0.2725$ & $0.7580$ & $0.3351$\\ 
\hline
$\log_{10}(\frac{M_{\ast}}{M_{\odot}})$ & $ 0.2681$ & $0.5357$ & $0.7697$ & $0.1029$ & $-0.1769$ & $0.0831$ \\
\hline
$\log_{10}(\frac{sSFR}{Gyr^{-1}})$ & $-0.4508$ & $0.3698$ & $0.1248$ & $-0.2731$ & $0.6156$ & $-0.4369$ \\
\hline
Metallicity & $0.2172$ & $0.7364$ & $-0.6258$ & $0.0701$ & $-0.1177$ & $0.0107$ \\
\hline
$\frac{r_{90}}{r_{50}}$ & $0.4464$ & $-0.0516$ & $-0.0056$ & $-0.8914$ & $0.0050$ & $0.0585$ \\
\hline
$D4000$ & $0.4945$ & $-0.1631$ & $-0.0049$ & $0.2026$ & $-0.0370$ & $-0.8285$ \\
\hline
\end{tabular}
\caption{The weights or coefficients of each original variable in the
  linear combinations that form the principal components are the
  elements of the corresponding eigenvector. This table shows the
  elements of the eigenvector obtained from the combined stellar
  mass-matched sample.}
\label{tab:pca}
\end{table*}

%%%%%%%%%%%%%%%%%%%%%%%%%%%%%%%%%%%%%%%%%%%%%%%%%%%%%%%%%%%%%%%%%%%%%%%%%%%%%%%%%%%
\begin{table}[htbp]
    \centering
    \begin{tabular}{|c|c|}
    \hline
     Principal Component & Percentage of Variance Explained \\
     \hline
       PC1  & $55.21$ \\
       \hline
       PC2 & $16.25$ \\
       \hline
       PC3 & $13.14$ \\
       \hline
       PC4 & $7.06$ \\
       \hline
       PC5 & $4.79$ \\
       \hline
       PC6 & $3.55$ \\
       \hline
    \end{tabular}
    \caption{This table represents the percentage of variance explained by individual principal component.}
    \label{tab:variance}
\end{table}

%\begin{figure}[htbp]
%\centering
%\includegraphics[width=.46\textwidth]{explained_variances.pdf}
%\caption{This figure shows the percentage of variance explained by the
%  different principal components.}
%\label{fig:exp_var}
%\end{figure}
%%%%%%%%%%%%%%%%%%%%%%%%%%%%%%%%%%%%%%%%%%%%%%%%%%%%%%%%%%%%%%%%%%%%%%%%%%%%%%%%%%%
%%%%%%%%%%%%%%%%%%%%%%%%%%%%%%%%%%%%%%%%%%%%%%%%%%%%%%%%%%%%%%%%%%%%%%%%%%%%%%%%%%%
\begin{figure*}[htbp]
\centering
  \subcaptionbox*{}[.46\linewidth]{%
    \includegraphics[width=\linewidth]{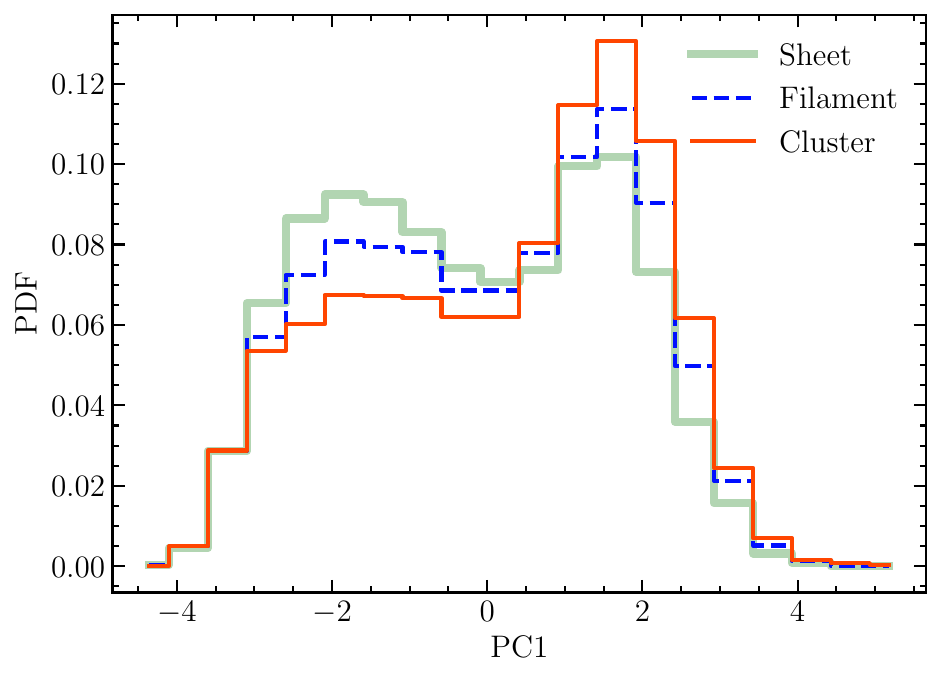}%
  }%
  \hfill
  \subcaptionbox*{}[.46\linewidth]{%
    \includegraphics[width=\linewidth]{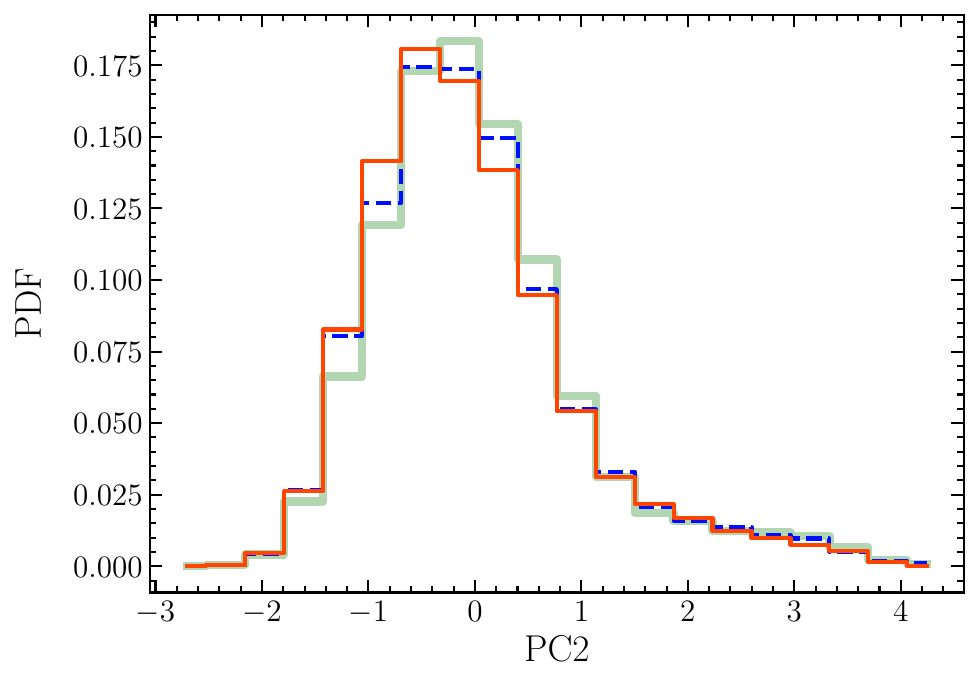}%
  }%
  \hfill
   \subcaptionbox*{}[.46\linewidth]{%
    \includegraphics[width=\linewidth]{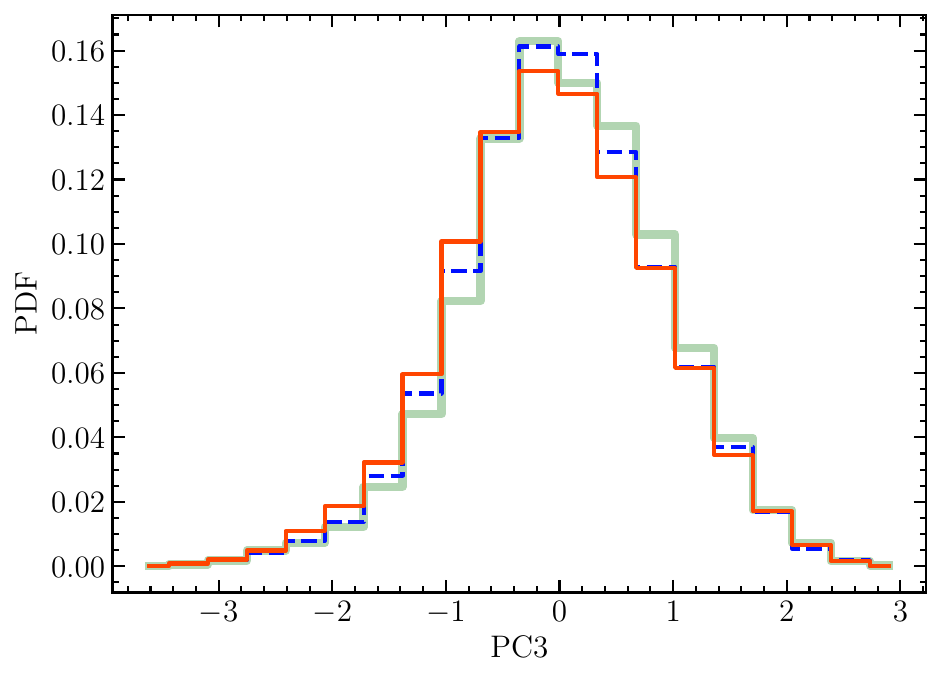}%
  }
  \caption{The top left, top right and bottom panels of this figure
    respectively compares the PDFs of PC1, PC2 and PC3 in different
    geometric environments.}
  \label{fig:pc_pdf}
\end{figure*}
%%%%%%%%%%%%%%%%%%%%%%%%%%%%%%%%%%%%%%%%%%%%%%%%%%%%%%%%%%%%%%%%%%%%%%%%%%%%%%%%%%%%%%%%%%%%%%%%%%%%%%%%%%%
%%%%%%%%%%%%%%%%%%%%%%%%%%%%%%%%%%%%%%%%%%%%%%%%%%%%%%%%%%%%%%%%%%%%%%%%%%%%%%%%%%%%%%%%%%%%%%%%%%%%%%%%%%%
\begin{table*}[htbp]
 \centering
 \begin{minipage}{.5\textwidth}
  \centering
    \begin{tabular}{|c|c|}
    \hline
    Web Environments & $p$-value \\
    \hline
    \hline
     Sheet-Filament & $8.22\times10^{-24}$ \\
     \hline
     Filament-Cluster & $5.67\times10^{-29}$ \\
     \hline
     Sheet-Cluster & $2.75\times 10^{-98}$ \\
    \hline
    \end{tabular}
    \caption{KS test results for PC1}
    \label{tab:kstest_pc1}
    \end{minipage}%
    \hfill
    \begin{minipage}{.5\textwidth}
    \centering  
    \begin{tabular}{|c|c|}
    \hline
    Web Environments & $p$-value \\
    \hline
    \hline
     Sheet-Filament & $9.42\times 10^{-6}$ \\
     \hline
     Filament-Cluster & $4.85\times 10^{-5}$ \\
     \hline
     Sheet-Cluster & $2.63\times10^{-17}$ \\
    \hline
    \end{tabular}
    \caption{KS test results for PC2}
    \label{tab:kstest_pc2}
    \end{minipage}%

    \vspace{1cm}  % Adjust the space here

    \begin{minipage}{.5\textwidth}
    \centering  
    \begin{tabular}{|c|c|}
    \hline
    Web Environments & $p$-value \\
    \hline
    \hline
     Sheet-Filament & $7.45\times 10^{-6}$ \\
     \hline
     Filament-Cluster & $9.71\times 10^{-8}$ \\
     \hline
     Sheet-Cluster & $2.30\times 10^{-18}$ \\
    \hline
    \end{tabular}
    \caption{KS test results for PC3}
    \label{tab:kstest_pc3}
    \end{minipage}
\end{table*}

%%%%%%%%%%%%%%%%%%%%%%%%%%%%%%%%%%%%%%%%%%%%%%%%%%%%%%%%%%%%%%%%%%%%%%%%%%%%%%%%%%%%%%%%%%%%
%%%%%%%%%%%%%%%%%%%%%%%%%%%%%%%%%%%%%%%%%%%%%%%%%%%%%%%%%%%%%%%%%%%%%%%%%%%%%%%%%%%%%%%%%%%%
\begin{figure*}[htbp]
\centering
\includegraphics[width=15cm]{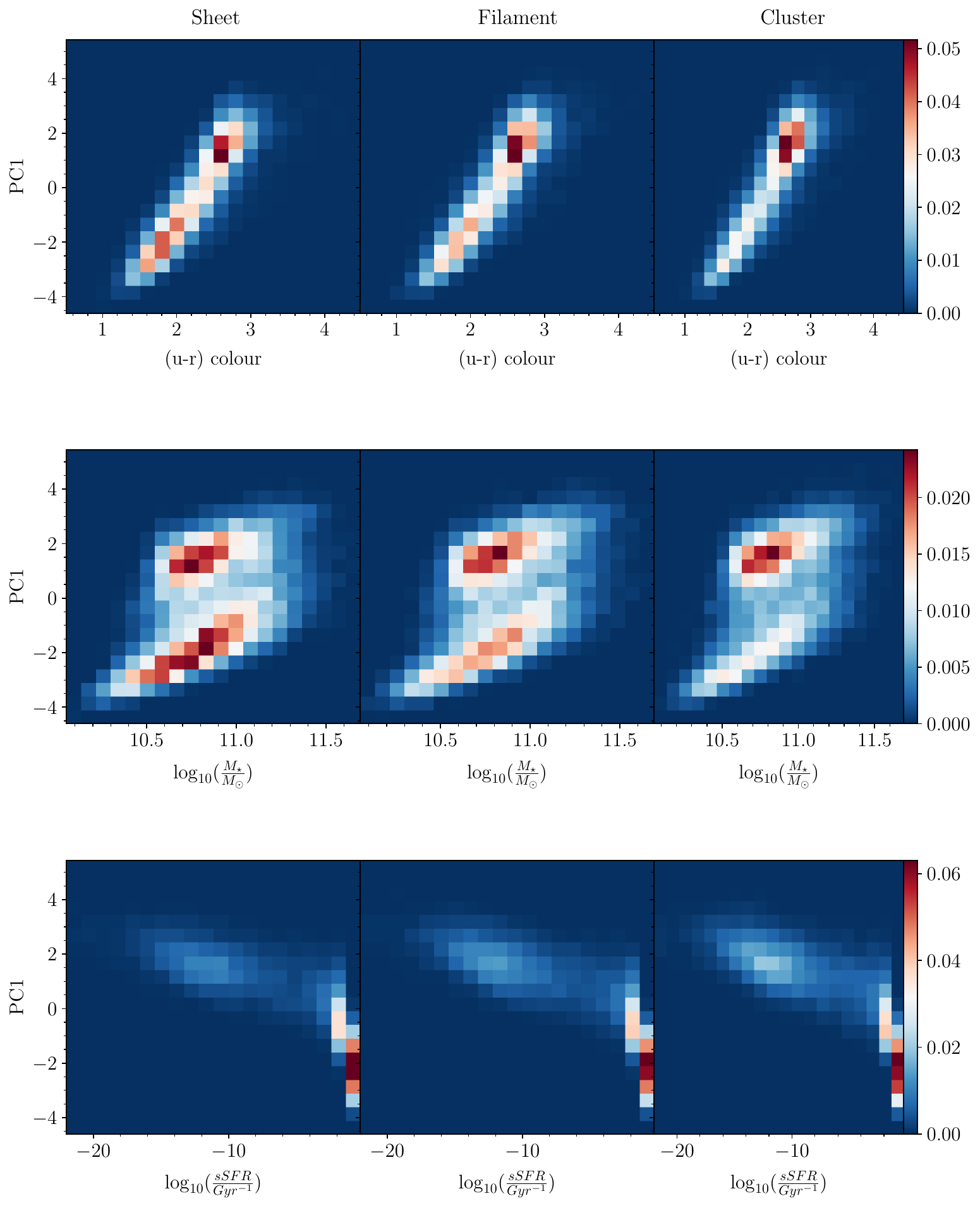}
\caption{The top panels of this figure show the joint PDFs of PC1 and
  $(u-r)$ colour across three cosmic web environments. The middle and
  lower panels show the same but for
  $\log_{10}(\frac{M_{\ast}}{M_{\odot}})$ and
  $\log_{10}(\frac{sSFR}{Gyr^{-1}})$ respectively.}
\label{fig:pc1_set1}
\end{figure*}
\newpage
\begin{figure*}[htbp]
\includegraphics[width=15cm]{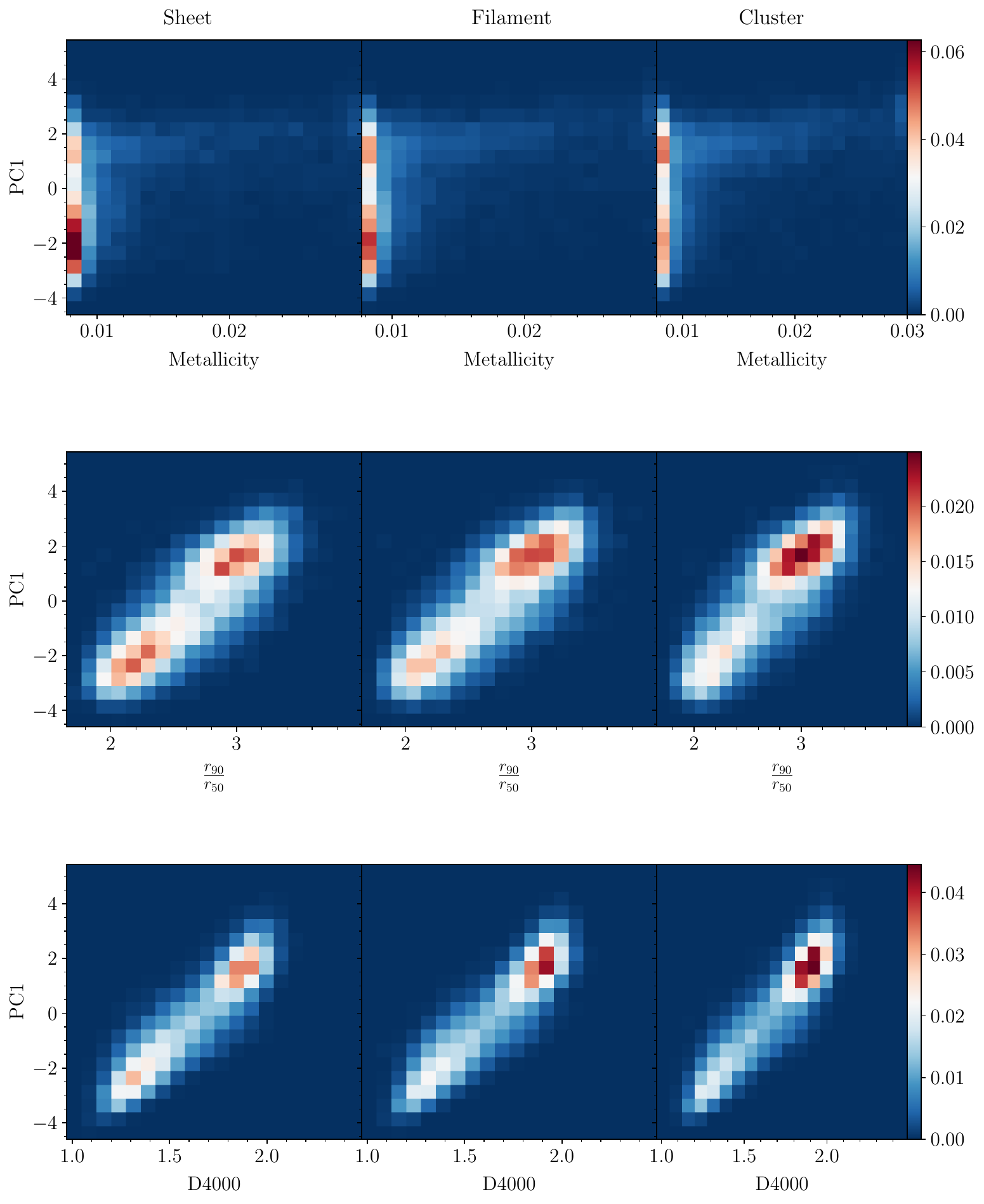}
\caption{Same as \autoref{fig:pc1_set1} but for metallicity,
  concentration index and D4000.}
\label{fig:pc1_set2}
\end{figure*}
%%%%%%%%%%%%%%%%%%%%%%%%%%%%%%%%%%%%%%%%%%%%%%%%%%%%%%%%%%%%%%%%%%%%%%%%%%%%%%%%%%%%%%%%%%%%%%%%%%%%%%%%%%%
\begin{figure*}[htbp]
\centering
\includegraphics[width=15cm]{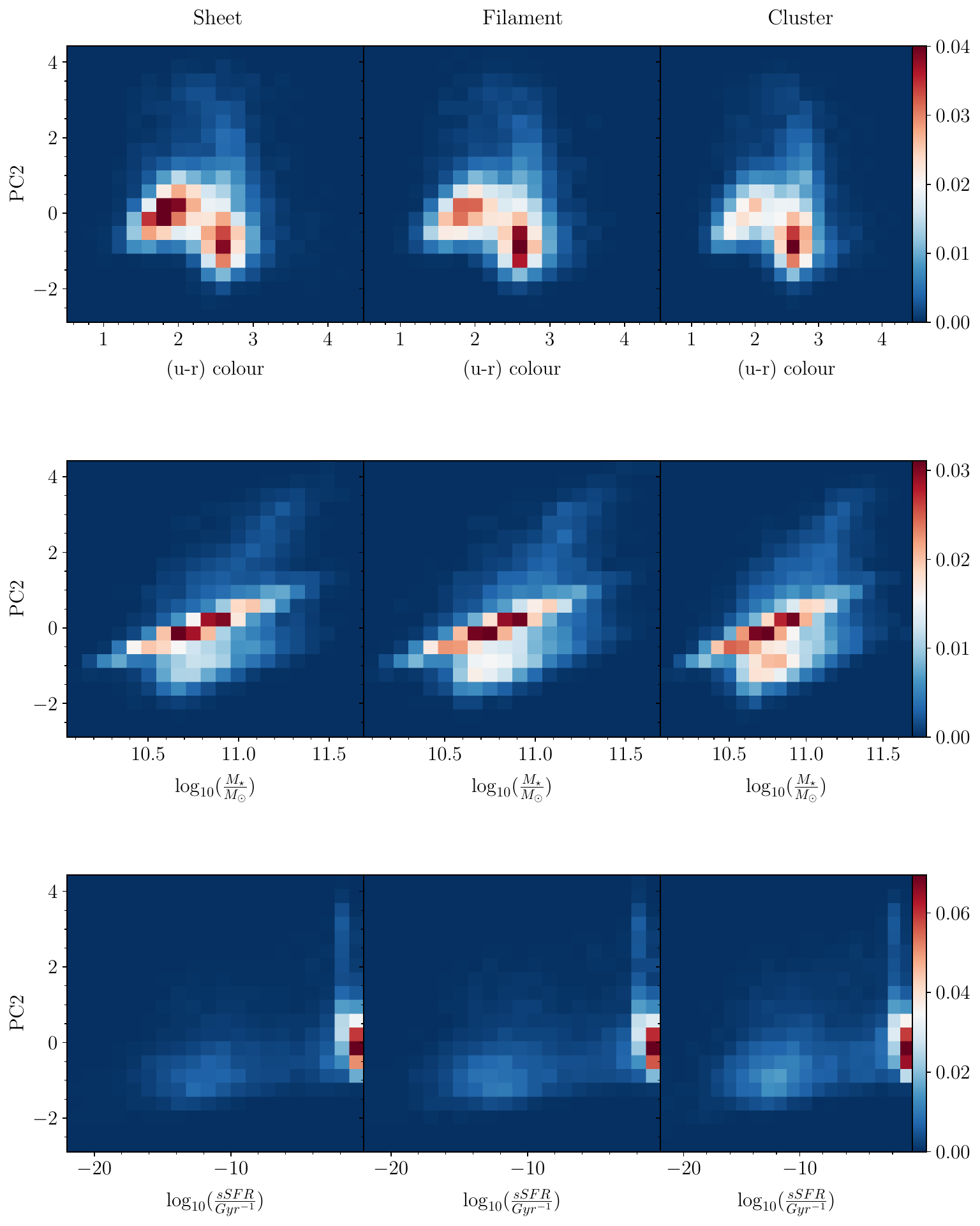}
\caption{Same as \autoref{fig:pc1_set1} but for PC2.}
\label{fig:pc2_set1}
\end{figure*}
\newpage
\begin{figure*}[htbp]
\includegraphics[width=15cm]{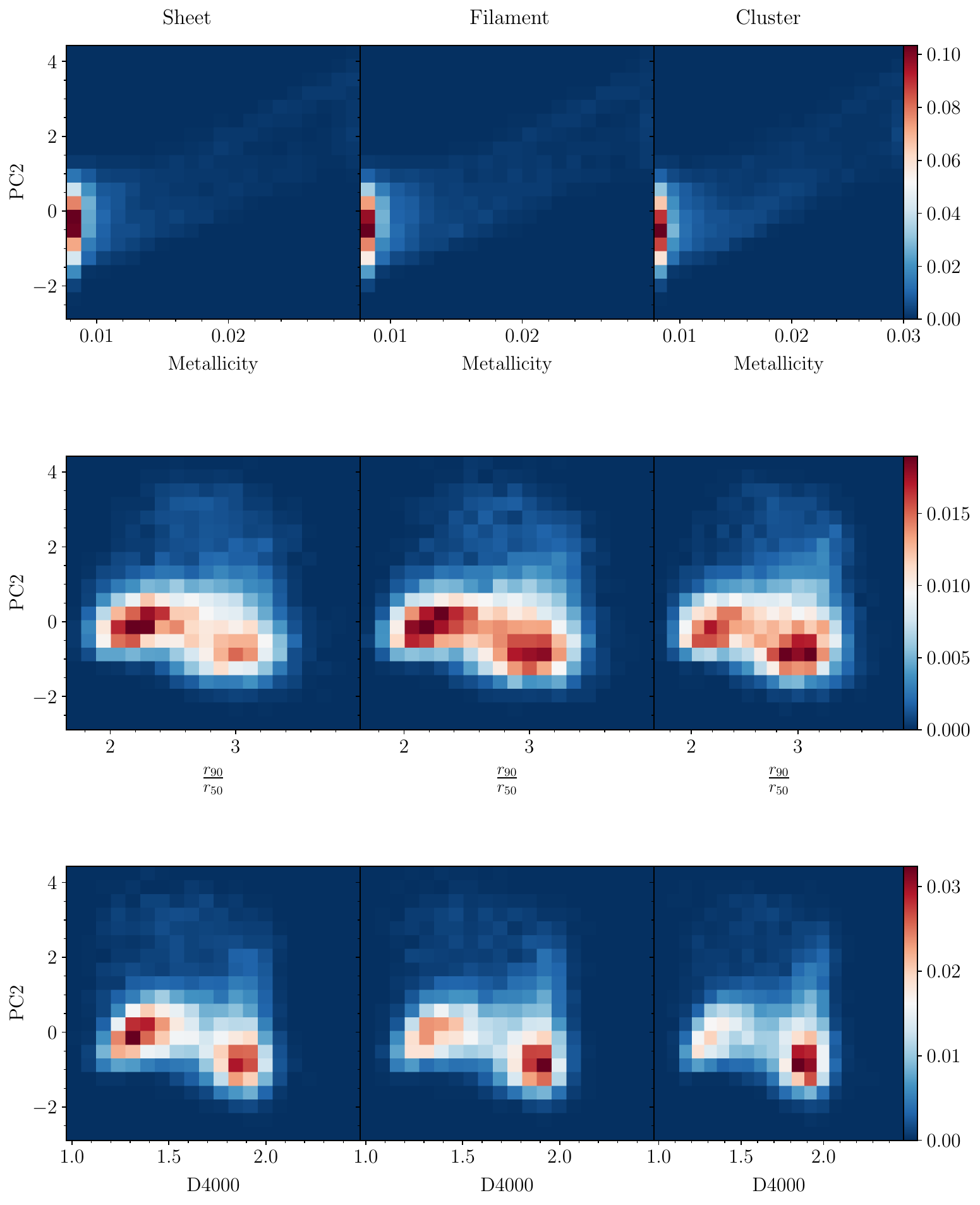}
\caption{Same as \autoref{fig:pc1_set2} but for PC2.}
\label{fig:pc2_set2}
\end{figure*}

%%%%%%%%%%%%%%%%%%%%%%%%%%%%%%%%%%%%%%%%%%%%%%%%%%%%%%%%%%%%%%%%%%%%%%%%%%%%%%%%%%%%%%%%%%%%%
%%%%%%%%%%%%%%%%%%%%%%%%%%%%%%%%%%%%%%%%%%%%%%%%%%%%%%%%%%%%%%%%%%%%%%%%%%%%%%%%%%%%%%%%%%%%%%%%%%%%%%%%%%%
\begin{figure*}[htbp]
\centering
\includegraphics[width=15cm]{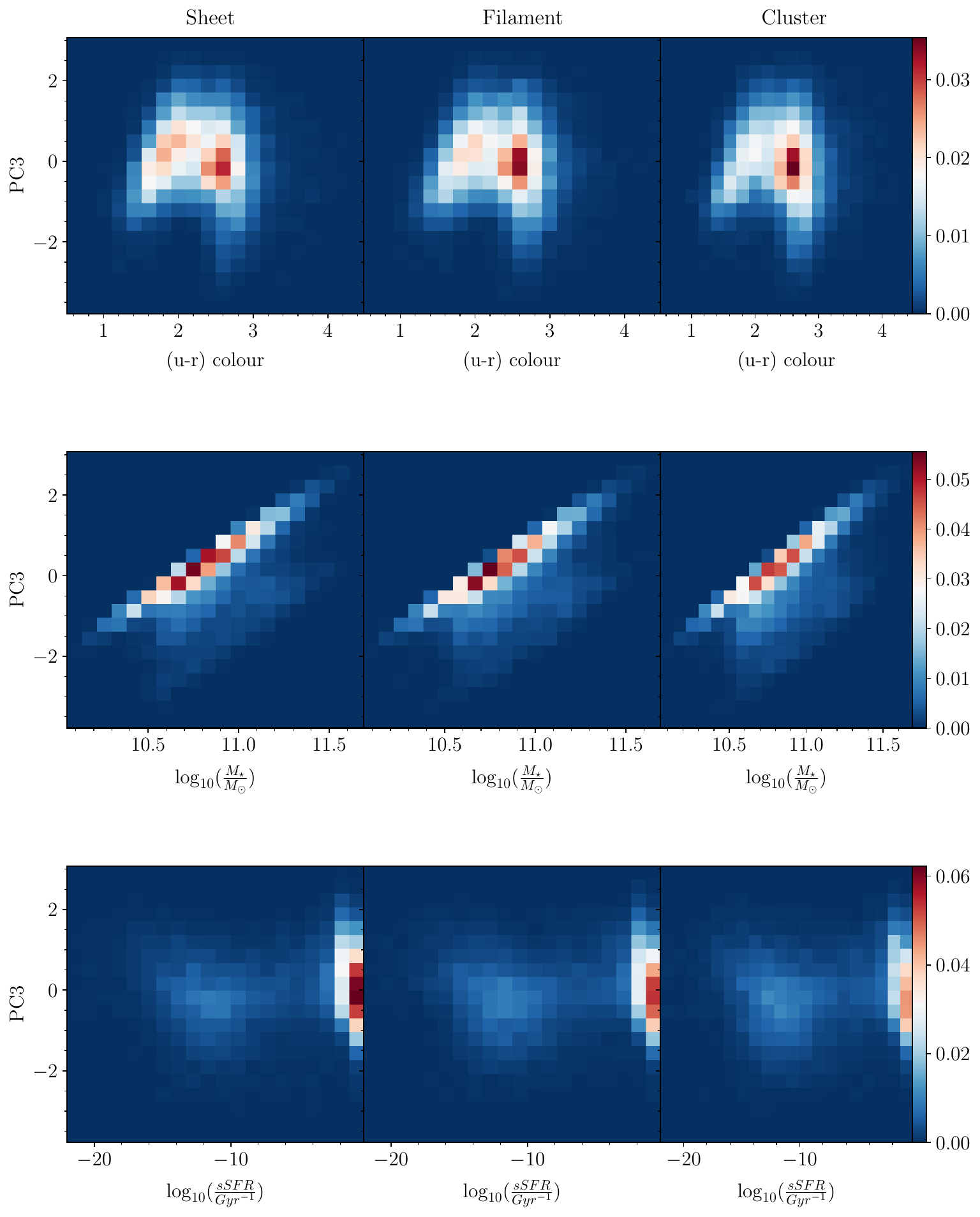}
\caption{Same as \autoref{fig:pc1_set1} but for PC3.}
\label{fig:pc3_set1}
\end{figure*}
\newpage
\begin{figure*}[htbp]
\includegraphics[width=15cm]{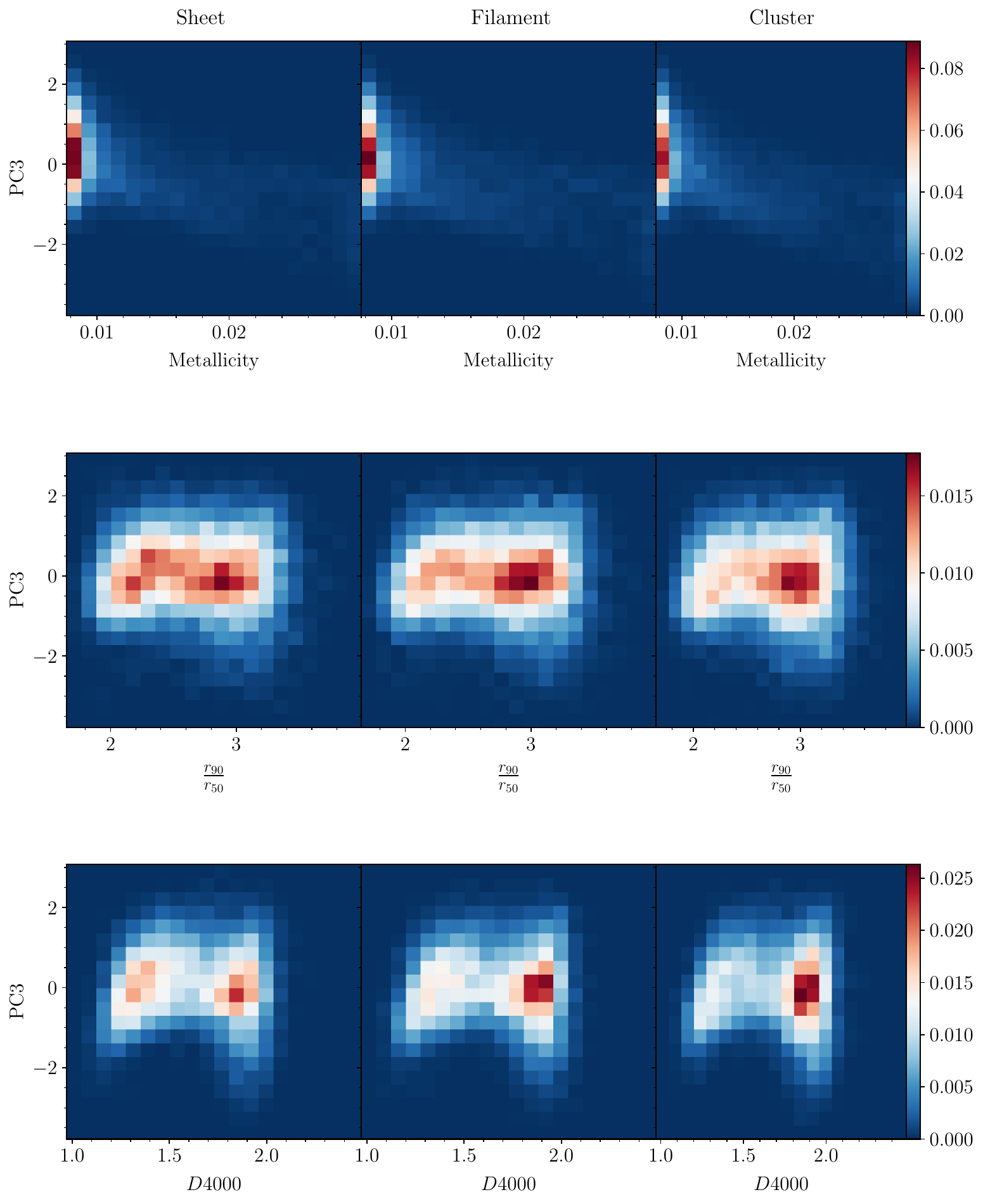}
\caption{Same as \autoref{fig:pc1_set2} but for PC3.}
\label{fig:pc3_set2}
\end{figure*}

%%%%%%%%%%%%%%%%%%%%%%%%%%%%%%%%%%%%%%%%%%%%%%%%%%%%%%%%%%%%%%%%%%%%%%%%%%%%%%%%%%%%%%%%%%%%%

\subsection{Comparing the principal components in different cosmic web environments}
\label{subsec:pca_web}
We categorize galaxies into various cosmic web environments and
compute PC1, PC2, and PC3 using eigenvectors derived from the entire
dataset for each environment. In the top left panel of
\autoref{fig:pc_pdf}, we compare the PDFs of PC1 across filaments,
sheets, and clusters. PC1 displays clear bimodality in all cosmic web
environments, with two distinct peaks symmetrically positioned around
PC $=0$. The amplitudes of these peaks differ across environments:
sheets exhibit the highest amplitude for the peak at negative PC1,
followed by filaments and clusters. The trend reverses for the peak at
positive PC1. The observed bimodality in the PDF of PC1 reflects
underlying structure where two distinct groups of galaxies are
present. This finding is consistent with the observed bimodality in
many galaxy properties like colour, sSFR, morphology and star
formation history. Interestingly, these four galaxy properties have
the dominant contribution to PC1 as compared to the other two
properties (stellar mass and metallicity). The symmetrical location of
the two peaks around PC1 $=0$ suggests that the two groups of galaxies
are roughly equal in size and have contrasting physical
characteristics.

PC1 encapsulates the interplay between sSFR, colour, morphology , and
D4000 in shaping galaxy bimodality, emphasizing how star formation and
aging processes co-evolve differently in different cosmic web
environments. The bimodality in PC1 reflects how environmental
processes act collectively to separate galaxies into star forming and
quenched groups.

In the top right panel of \autoref{fig:pc_pdf}, we compare the PDFs of
PC2 across different environments. PC2 is dominated by metallicity and
exhibit a unimodal distribution across all environments, indicating
less variability compared to PC1. Interestingly, the PDFs of PC2 in
each environment is asymmetrical around PC2 $=0$ and has an extended
tail towards the positive PC2 values. This aligns with the
mass-metallicity relation, where lower-mass, more numerous galaxies
tend to have lower metallicities, while high-metallicity galaxies are
rarer and more massive. The positive skewness in PC2 possibly also
captures subtler environmental influences on secondary processes like
chemical enrichment.

We compare the PDFs of PC3 across different environments in the bottom
panel of \autoref{fig:pc_pdf}. The distributions are unimodal in each
environment. However, they exhibit an extended tail towards the
negative PC3 values. PC3 is dominated by stellar mass. The negatively
skewed distribution indicates a concentration of high-mass galaxies
and a tail toward low masses. It also suggests that lower-mass
galaxies contribute less distinctive variation once PC1 and PC2 are
accounted for.

Thus PC1 highlights clear differences among galaxy properties in
different cosmic web environments whereas PC2, PC3 capture a more
continuous variation in physical properties with a positive
skewness. We assess the statistical significance of the differences in
PDFs of PC1, PC2 and PC3 across various cosmic web environment. The
results of the KS tests are detailed in \autoref{tab:kstest_pc1},
\autoref{tab:kstest_pc2}, and \autoref{tab:kstest_pc3}
respectively. The tables show that the null hypothesis can be rejected
at $>99.99\%$ confidence level in all cases suggesting that the first
three principal components are distinctly different in various
geometric environments of the cosmic web. The fact that PC1, PC2, PC3
behave differently across cosmic web environments implies that the
spatial arrangement of galaxies in sheets, filaments, or clusters have
a measurable impact on their physical properties.

%%%%%%%%%%%%%%%%%%%%%%%%%%%%%%%%%%%%%%%%%%%%%%%%%%%%%%%%%%%%%%%%%%%%%%%%%%%%%%%%%%%%%%%%%%%%%
\subsection{Comparing the correlations between principal components and galaxy properties in different geometric environments}
\label{sec:pca_prop}
Principal components are linear combinations of various galaxy
properties, so their relationships might be approximately
linear. However, no galaxy property is perfectly correlated with the
principal components. The complex interplay between different galaxy
properties and their varying contributions to the principal components
indicates that modeling principal components as a linear function of
any single galaxy property would capture only part of the overall
picture. Generally, the relationship can be non-linear if other galaxy
properties significantly influence the principal components. With this
in mind, we use normalized mutual information (NMI) to explore the
individual relationships between each principal component and galaxy
property.

We show the joint probability distributions of PC1 with the six
different galaxy properties across various geometric environments in
\autoref{fig:pc1_set1} and \autoref{fig:pc1_set2}. In each
environment, the various panels of these figures indicate that the
$(u-r)$ colour, $\frac{r_{90}}{r_{50}}$, and D4000 show positive
correlations with PC1. Conversely, $\log_{10}(\frac{sSFR}{Gyr^{-1}})$
exhibits a negative correlation with PC1 in all environments. It may
be noted here that NMI does provide any direction of the correlations
unlike Pearson correlation coefficient. However, one can find the
general trends by inspecting the sign of the corresponding element of
the eigen vector (representing PC1) or analyzing the joint PDF of PC1
and a galxy property. The stellar mass and metallicity exhibit a
positive correlation with PC1. However their relationships seems to be
more complex. These trends are consistent with the sign and magnitude
of the elements of the first eigenvector $v_{1}$ showm in
Table~\ref{tab:pca}. The various panels in \autoref{fig:pc1_set1} and
\autoref{fig:pc1_set2} demonstrate that the joint PDFs of PC1 and
different galaxy properties are sensitive to the geometric
environments. It may be noted that we use the exact same number of
mass-matched galaxies from each cosmic web environment.

It is well established that galaxies rich in gas typically exhibit
higher star formation rates, bluer colours, younger stellar
populations, and disk-like morphologies. Our analysis reveals that the
correlations between PC1 and properties related to star formation, gas
content, or morphology vary significantly across different cosmic web
environments. This variability reflects how galaxy properties are
influenced by factors such as gas accessibility, galaxy interactions,
the effectiveness of various quenching mechanisms, and tidal
interactions. Untangling the individual contributions of each of these
factors to the observed differences in the joint distributions is
extremely challenging. We present the joint PDFs for PC2 and different
galaxy properties in \autoref{fig:pc2_set1} and
\autoref{fig:pc2_set2}. Similarly, the joint PDFs of PC3 and different
galaxy properties are shown in \autoref{fig:pc3_set1} and
\autoref{fig:pc3_set2}.These figures reveal that the shapes and
spreads of the joint PDFs are markedly influenced by the cosmic web
environments. These variations underscore the significant impact of
geometric environments on shaping galaxy properties and their
interrelationships.

We measure the NMI between the principal components (PC1, PC2 and PC3)
and the individual galaxy properties, and then assess the statistical
significance of the differences in its value across different cosmic
web environments. To do this, we generate 50 jackknife samples from
the original data by randomly sampling $80\%$ of the galaxies each
time, without replacement. We estimate the principal components from
each of these 50 jackknife samples and calculate the normalized mutual
information (NMI) (\autoref{sec:method3}) between the principal
components and the individual galaxy properties. We then use a
two-tailed t-test to determine if there is a significant difference
between the mean NMI values across different pairs of geometric
environments. The results of our test for PC1, PC2 and PC3 are
presented in \autoref{tab:pc1ttest}, \autoref{tab:pc2ttest}, and
\autoref{tab:pc3ttest} respectively. We observe that the p-value for
each relationship in every pair of geometric environments is extremely
small ($<10^{-4}$ in nearly all cases), indicating that the measured
NMI for any relationship is significantly different ($>99.99\%$
confidence level) across various cosmic web environments. The very
small p-values observed in our analysis are likely a consequence of
the large sample size and the increased statistical power it
provides. These small values indicate that the correlations detected
between principal components and galaxy properties across different
cosmic web environments are statistically robust and unlikely to have
arisen by chance. However, while the statistical significance is
evident, the physical significance requires careful interpretation. In
the context of galaxy evolution in the cosmic web, even weak
correlations can be physically meaningful if they point to consistent
environmental trends.  These subtle trends can reveal important clues
about mechanisms like gas accretion, quenching, or galaxy
interactions, which operate differently depending on the large-scale
cosmic web environment. Therefore, identifying and interpreting even
modest correlations can be crucial for understanding how the cosmic
web influences the formation and evolution of galaxies. Our analysis
suggests that the cosmic web environment has a significant influence
on galaxy evolution. PCA shows that galaxies in different cosmic web
environments differ in complex ways that are not evident from the
correlations between individual galaxy properties.

We perform these analyses using $20$ bins for each
variable. Additionally, we repeat the analysis with $10$ and $30$ bins
and find that our main conclusions remain unchanged. We also repeat
the entire analysis using the dataset without specifying any ranges
for the galaxy properties and find that our conclusions remain the
same.

\begin{table}
\centering
\begin{tabular}{|c|c|c|c|c|}
\hline
{\rule{0pt}{4ex} Relations} &  Sheet-Filament  & Filament-Cluster & Sheet-Cluster  \\ \cline{2-4} 
 &  $p$ value & $p$ value & $p$ value \\
\hline\hline
\rule{0pt}{3ex} PC1 - colour  & $9.09\times 10^{-60}$ & $2.11\times 10^{-28}$  & $3.90\times 10^{-73}$ \\
\hline
\rule{0pt}{3ex} PC1 - stellar mass & $9.45\times 10^{-9}$  &$2.55 \times 10^{-17}$  & $2.79 \times 10^{-4}$  \\
\hline
\rule{0pt}{3ex} PC1 - sSFR & $3.97\times10^{-59}$ & $2.26\times 10^{-67}$  & $3.93\times 10^{-88}$ \\
\hline
\rule{0pt}{3ex} PC1 - metallicity  & $4.69\times 10^{-9}$ & $5.44 \times 10^{-90}$  & $6.00\times 10^{-89}$  \\
\hline
\rule{0pt}{3ex} PC1 - concentration index & $2.14\times 10^{-7}$ & $5.61\times 10^{-38}$ & $5.79\times 10^{-46}$  \\
\hline
\rule{0pt}{3ex} PC1 - $D4000$ & $5.80\times 10^{-52}$ & $5.26\times 10^{-50}$ & $5.90 \times 10^{-79}$ \\
\hline
\end{tabular}
\caption{This table summarizes the results of a two-tailed t-test
  after comparing the normalized mutual information for the NMI
  between PC1 and different galaxy properties in two different cosmic
  web environments. The degrees of freedom in this test is $(n_1 + n_2
  -2) = 98$.}
\label{tab:pc1ttest}
\end{table}

%%%%%%%%%%%%%%%%%%%%%%%%%%%%%%%%%%%%%%%%%%%%%%%%%%%%%%%%%%%%%%%%%%%%%%%%%%
%%%%%%%%%%%%%%%%%%%%%%%%%%%%%%%%%%%%%%%%%%%%%%%%%%%%%%%%%%%%%%%%%%%%%%%%%%

\begin{table}
\centering
\begin{tabular}{|c|c|c|c|c|}
\hline
{\rule{0pt}{4ex} Relations} &  Sheet-Filament  & Filament-Cluster & Sheet-Cluster  \\ \cline{2-4} 
 &  $p$ value & $p$ value & $p$ value \\
\hline\hline
\rule{0pt}{3ex} PC2 - colour  & $2.64 \times 10^{-55}$ & $6.19 \times 10^{-39}$  & $4.97\times 10^{-75}$ \\
\hline
\rule{0pt}{3ex} PC2 - stellar mass & $3.16 \times 10^{-69}$  &$3.32 \times 10^{-65}$  & $4.12 \times 10^{-96}$  \\
\hline
\rule{0pt}{3ex} PC2 - sSFR & $2.81 \times 10^{-34}$ & $3.99 \times 10^{-64}$  & $1.83\times 10^{-77}$ \\
\hline
\rule{0pt}{3ex} PC2 - metallicity  & $8.05 \times 10^{-41}$ & $4.01 \times 10^{-16}$  & $4.05 \times 10^{-57}$  \\
\hline
\rule{0pt}{3ex} PC2 - concentration index & $5.45 \times 10^{-18}$ & $7.35 \times 10^{-43}$ & $3.26\times 10^{-53}$  \\
\hline
\rule{0pt}{3ex} PC2 - $D4000$ & $2.77\times 10^{-20}$ & $2.81 \times 10^{-30}$ & $2.09\times 10^{-51}$ \\
\hline
\end{tabular}
\caption{Same as \autoref{tab:pc1ttest}, but for the NMI between PC2 and different galaxy properties.}
\label{tab:pc2ttest}
\end{table}

%%%%%%%%%%%%%%%%%%%%%%%%%%%%%%%%%%%%%%%%%%%%%%%%%%%%%%%%%%%%%%%%%%%%%%%%%%%%%%%%%%%%%%%%%%%%%

\begin{table}[h!]
\centering
\begin{tabular}{|c|c|c|c|c|}
\hline
{\rule{0pt}{4ex} Relations} &  Sheet-Filament & Filament-Cluster & Sheet-Cluster  \\ \cline{2-4} 
 &  $p$ value & $p$ value & $p$ value \\
\hline\hline
\rule{0pt}{3ex} PC3 - colour & $5.71 \times 10^{-63}$ & $3.10 \times 10^{-6}$ & $2.65 \times 10^{-60}$ \\
\hline
\rule{0pt}{3ex} PC3 - stellar mass & $1.35 \times 10^{-74}$ & $3.19 \times 10^{-40}$ & $2.29 \times 10^{-87}$ \\
\hline
\rule{0pt}{3ex} PC3 - sSFR & $1.27 \times 10^{-14}$ & $4.71 \times 10^{-19}$ & $2.72 \times 10^{-35}$ \\
\hline
\rule{0pt}{3ex} PC3 - metallicity & $1.38 \times 10^{-3}$ & $1.32 \times 10^{-66}$ & $1.62 \times 10^{-60}$ \\
\hline
\rule{0pt}{3ex} PC3 - concentration index & $0.14$ & $9.35 \times 10^{-5}$ & $0.01$ \\
\hline
\rule{0pt}{3ex} PC3 - $D4000$ & $6.57 \times 10^{-55}$ & $1.52 \times 10^{-37}$ & $1.60 \times 10^{-21}$ \\
\hline
\hline
\end{tabular}
\caption{Same as \autoref{tab:pc1ttest}, but for the NMI between PC3 and different galaxy properties.}
\label{tab:pc3ttest}
\end{table}

%%%%%%%%%%%%%%%%%%%%%%%%%%%%%%%%%%%%%%%%%%%%%%%%%%%%%%%%%%%%%%%%%%%%%%%%%%

\section{Discussions and Conclusions}
In this work, we use Principal Component Analysis (PCA) to investigate
the influence of cosmic web environments (sheets, filaments, and
clusters) on galaxy properties, focusing on the distributions of the
principal components (PC1, PC2, and PC3), and their correlations with
physical properties such as colour, sSFR, D4000, metallicity, and
morphology. The results highlight the interplay between external
environmental effects and internal secular processes in shaping galaxy
evolution.

We analyze the principal components in different cosmic web
environments and test if their probability distributions are different
in a statistically significant manner. We also study the relationships
between the different principal components and individual galaxy
properties using the normalized mutual information (NMI). Our key
findings are summarized as follows.

PC1, PC2, and PC3 respectively explains $\sim 55\%$, $\sim 16\%$, and
$\sim 13\%$ variance in our data. PC1 shows clear bimodality in all
cosmic web environments. The presence of bimodality in PC1 indicates
its ability to account for most of the variance in the data. Two peaks
in PC1 are symmetrically located around PC1 $=0$. The nearly symmetric
placement of the two peaks in all environments suggests that two
groups of galaxies with contrasting physical properties are present in
each cosmic web environment. It implies that bimodality in different
galaxy properties can emerge in each type of geometric environment. We
note that the negative PC1 peak is highest in sheets, followed by
filaments, and lowest in clusters. The positive PC1 peak shows the
opposite trend, highest in clusters, followed by filaments, and lowest
in sheets. Thus the cosmic web environments strongly influence the
distribution of PC1, with sheets favouring negative PC1 and clusters
favouring positive PC1. PC1 is strongly correlated with properties
directly tied to star formation activity, including colour, sSFR,
D4000, and morphology. Its bimodal distribution reflects the two
dominant galaxy populations: star-forming galaxies (negative PC1
values) and quenched galaxies (positive PC1 values). Dividing galaxies
into two groups based on whether their PC1 values are negative or
positive could be an effective method for distinguishing quenched
galaxies from star-forming ones. This classification is particularly
advantageous because PC1 is a linear combination of several galaxy
properties, and it accounts for the bimodality observed in each of
these properties, making it a more robust approach.

We note that the bimodality in PC1 is strongly sensitive to the cosmic
web environment. In sheets, galaxies are predominantly star-forming,
maintaining blue colours, high sSFR, disk-like morphology and younger
stellar populations due to mild gravitational interactions and access
to fresh gas. In clusters, by contrast, galaxies are more likely to be
quenched, with redder colours, low sSFR, bulge-dominated morphology,
and older stellar populations. This reflects the effectiveness of
environmental mechanisms such as ram-pressure stripping, mergers, and
tidal interactions in suppressing star formation.

On the other hand, PC2 and PC3 do not exhibit bimodality in any cosmic
web environment. They exhibit less variability compared to PC1 across
cosmic web environments, implying they are less influenced by spatial
clustering. PC2 is dominated by metallicity and secondary variations
in stellar mass and morphology. Unlike PC1, PC2 exhibits a unimodal
and positively skewed distribution across all environments, suggesting
that metallicity evolves more gradually and is less sensitive to
sudden environmental changes. The weaker environmental dependence of
PC2 highlights the role of internal processes, such as gradual
chemical enrichment and mass growth through accretion, that proceed
independently of external conditions. On the other hand, PC3 exhibit a
unimodal distribution with negative skewness. It is dominated by
stellar mass, implying most of the variance explained by this
component comes from differences in stellar mass between galaxies
after accounting for the effects captured in PC1 and PC2. It may also
arise due to the fact that our volume limited sample have galaxies
above a certain mass. So the low-mass galaxies will be
underrepresented, skewing the distribution.

Using a KS test, we find that the null hypothesis can be rejected at
$>99.99\%$ confidence level for PC1, PC2, PC3. This suggests that
cosmic web plays a crucial role in shaping the correlations between
galaxy properties. The contrasting behavior of PC1, PC2, and PC3
encapsulates the balance between external and internal processes
shaping galaxies. PC1 is highly sensitive to environment and reflects
the immediate impact of the cosmic web on star formation and
quenching. PC2 and PC3, by comparison, are less environment-dependent
and captures longer-term, gradual processes like chemical enrichment
and stellar mass growth. The use of PCA thus provides a powerful
framework for understanding galaxy evolution, allowing us to capture
both dominant environmental effects and subtler secular trends.

We measure the normalized mutual information between principal
components and individual galaxy properties in different geometric
environments of the cosmic web. We use a two-tailed t-test to compare
the normalized mutual information in each pair of geometric
environment. We find that for each relationship and each pair of
geometric environment, the null hypothesis can be rejected at
$>99.99\%$ confidence level.

Understanding these differences is crucial as they provide insights
into how galaxies evolve and interact within their larger-scale
geometric environments. It also helps refine models of galaxy
formation and evolution. The cosmic web serves as a grand scaffolding
that intricately organizes and molds the distribution of matter and
galaxies across the vast expanse of the universe. Our analysis
suggests that various morphological components of the cosmic web like
filaments, sheets, clusters and voids have a significant impact on the
galaxy properties and their interrelationships. The galaxies in
clusters might evolve differently compared to those in filaments or
sheets, affecting their observed properties and correlations.  For
instance, filaments are the largest known coherent structures
\citep{pandey11, sarkar23} in the cosmic web, containing a significant
amount of baryonic matter, including WHIM
\citep{nicastro18}. Filaments host more than $80\%$ of the WHIM in the
universe \citep{tuominen21, galarraga21}. The continuous supply of gas
can influence the star formation rates, chemical enrichment, and
overall evolution of galaxies residing within filaments. Further,
studies with high-resolution zoom-in hydrodynamical simulations by
\cite{liao19} show that gas accretion onto filament halos is highly
anisotropic, leading to higher baryon and stellar fractions compared
to galaxies in the field. The low-mass halos typically accrete matter
perpendicular to their host filament, while high-mass halos mainly
accrete along the axis of the filament \citep{veena18}. Numerous
studies based on observational and simulations \citep{kuutma17,
  malavasi17, laigle18, kraljic18, bonjean21} have consistently shown
that galaxies near filament spines are redder, more massive, and less
star-forming than their counterparts, even at fixed environmental
density. This trend is commonly attributed to enhanced gas accretion,
more frequent galaxy-galaxy interactions, and increased merger rates
in these regions, which collectively drive rapid stellar mass
growth. Cosmological and hydrodynamical simulations \citep{singh20,
  song21, malavasi22, kotecha22} lend strong support to this
interpretation. These differences could be reflected in the PCA
results as distinct patterns or correlations between the principal
components and galaxy properties. The strong correlations of PC1 with
colour, sSFR, morphology, and D4000 can be interpreted as evidence
that environmental processes drive star formation and aging. The
bimodality in PC1 provides direct evidence of environmental effects,
with negative PC1 values reflecting sustained star formation in sheets
and filaments, and positive PC1 values reflecting quenching in
clusters. The positive skewness of PC2, the negative skewness of PC3
and their weaker environmental dependence suggest that properties like
metallicity and stellar mass evolve more gradually across different
cosmic web environment.

It is important to acknowledge some caveats in our analysis. We
conduct our study in redshift space, where the ``Finger of God''
effect can introduce some spurious filaments in dense groups and
clusters. The FoG effects primarily distort structures on smaller
scales (e.g., groups and some clusters), while the large-scale density
field remains relatively unaffected. We smooth the desnity field on
grids using a Gaussian Filter of width 8 Mpc. This implies that our
study focuses on large-scale environments where the impact of FoGs
should be minimal. Our volume limited sample also do not contain very
high density groups and clusters due to the spectroscopic
incompletenes arising from the finite size of the SDSS fibres and the
fibre collisions. Our main emphasis lies in exploring the impact of
the large-scale geometric environments in the cosmic web on the
interrelationships between various galaxy properties. We do not expect
FoGs to bias our measurements of the large-scale environment due to
these reasons. Further, the observed trends in PC1, PC2 and PC3 might
reflect a combination of local and global influences, as local density
naturally correlates with the global cosmic web structure. So the
differences observed in our analysis is a combined outcome of the
local density of galaxies and the large-scale cosmic web
environment. We tried to repeat our analysis by simulataneously
matching the stellar mass and the local density in all types of cosmic
web environments. Unfortunately, this provides us with a small number
of galaxies which represent only the lowest density regions from each
type of cosmic web environment.

Finally, our study demonstrates that PCA provides a powerful tool for
dissecting the impacts of the cosmic web on galaxy evolution. The
results highlight the complementary roles of external environmental
effects and internal secular processes in shaping galaxy properties
and their correlations.

\section*{ACKNOWLEDGEMENT}
Authors thank the anonymous reviewers for the valuable comments and
suggestions that helped us to improve the draft. AN acknowledges the
financial support from the Department of Science and Technology (DST),
Government of India through an INSPIRE fellowship. BP would like to
acknowledge financial support from the SERB, DST, Government of India
through the project CRG/2019/001110. BP would also like to acknowledge
IUCAA, Pune, for providing support through the associateship
programme.

We would also like to acknowledge the use of several Python packages such
as Pandas \citep{mckinney10}, NumPy \citep{harris20}, SciPy
\citep{virtanen20} and Matplotlib \citep{hunter07} in our work.

Funding for the SDSS and SDSS-II has been provided by the Alfred
P. Sloan Foundation, the Participating Institutions, the National
Science Foundation, the U.S. Department of Energy, the National
Aeronautics and Space Administration, the Japanese Monbukagakusho, the
Max Planck Society, and the Higher Education Funding Council for
England. The SDSS website is http://www.sdss.org/.

The SDSS is managed by the Astrophysical Research Consortium for the
Participating Institutions. The Participating Institutions are the
American Museum of Natural History, Astrophysical Institute Potsdam,
University of Basel, University of Cambridge, Case Western Reserve
University, University of Chicago, Drexel University, Fermilab, the
Institute for Advanced Study, the Japan Participation Group, Johns
Hopkins University, the Joint Institute for Nuclear Astrophysics, the
Kavli Institute for Particle Astrophysics and Cosmology, the Korean
Scientist Group, the Chinese Academy of Sciences (LAMOST), Los Alamos
National Laboratory, the Max-Planck-Institute for Astronomy (MPIA),
the Max-Planck-Institute for Astrophysics (MPA), New Mexico State
University, Ohio State University, University of Pittsburgh,
University of Portsmouth, Princeton University, the United States
Naval Observatory, and the University of Washington.

%%\appendix

%%\section{Appendix title 1}

%%\section{Appendix title 2}

%%\bibliographystyle{elsarticle-harv}

%%\bibliographystyle{elsarticle-num}
\bibliographystyle{unsrtnat}
% \bibliography{example}
\bibliography{ref}



%% For example:

%% \bibitem[Aladro et al.(2015)]{Aladro15} Aladro, R., Martín, S., Riquelme, D., et al. 2015, \aas, 579, A101

%%\end{thebibliography}

\end{document}